\documentclass[
  aps,
  prd,
  twocolumn,
  10pt,
  amsmath,
  amssymb,
  nofootinbib,
  superscriptaddress,
  floatfix,
]{revtex4-2}

\usepackage{graphicx}
\usepackage{booktabs}
\usepackage{xcolor}
\usepackage{hyperref}
\hypersetup{colorlinks=true, linkcolor=blue, citecolor=blue, urlcolor=blue}

\graphicspath{{figures/}}

\newcommand{\vect}[1]{\boldsymbol{#1}}
\newcommand{\psiBL}{\psi_{\mathrm{BL}}}
\newcommand{\Ahat}{\hat{A}}
\newcommand{\dID}{\partial \mathrm{ID}/\partial\theta}
\newcommand{\Madm}{M_{\mathrm{ADM}}}

\begin{document}

\title{A Differentiable Parametric Model of Binary-Black-Hole Initial Data: \texorpdfstring{\\}{ }I. Conformally flat Bowen--York punctures}

\author{Frederik De Ceuster}
\email{frederik.deceuster@kuleuven.be}
\affiliation{Institute for Theoretical Physics, KU Leuven, Celestijnenlaan 200D, 3001 Leuven, Belgium}
\affiliation{Leuven Gravity Institute, KU Leuven, Celestijnenlaan 200D, 3001 Leuven, Belgium}
\affiliation{STADIUS Center for Dynamical Systems, Signal Processing and Data Analytics, KU Leuven, Kasteelpark Arenberg 10, 3001 Leuven, Belgium}

\author{Tjonnie G.-F. Li}
\affiliation{Institute for Theoretical Physics, KU Leuven, Celestijnenlaan 200D, 3001 Leuven, Belgium}
\affiliation{Leuven Gravity Institute, KU Leuven, Celestijnenlaan 200D, 3001 Leuven, Belgium}
\affiliation{STADIUS Center for Dynamical Systems, Signal Processing and Data Analytics, KU Leuven, Kasteelpark Arenberg 10, 3001 Leuven, Belgium}

\date{\today}

\begin{abstract}
Constraint-satisfying initial data for binary black holes require the repeated solution of a nonlinear elliptic boundary-value problem throughout the physical parameter space.
We present a para\-metric, differentiable initial-data model that combines sparse-grid spectral interpolation with Newton refinement over a family of binary configurations.
Interpolation is used only to provide warm starts for the underlying elliptic solver, rather than to replace it.
A small number of Newton iterations refines each interpolated solution to a discrete, row-equilibrated constraint residual $\|R\|_\infty\le10^{-10}$, independent of the interpolation error, so that every returned initial-data set satisfies the discretized constraint to that tolerance.
Our JAX implementation provides analytic derivatives with respect to the physical parameters, enabling gradient-based parameter targeting, optimization, and sensitivity analysis.
We demonstrate the method for quasi-circular binary-black-hole initial data using conformally flat Bowen--York punctures by constructing four-dimensional aligned-spin and eight-dimensional general-spin models.
The resulting initial data are validated against TwoPunctures, and the method is applied to certified targeting of the physical Arnowitt--Deser--Misner mass and angular momentum and to effective-potential eccentricity reduction using a dedicated separation--tangential-momentum model.
\end{abstract}

\maketitle

\section{Introduction}
\label{sec:intro}
Every numerical-relativity simulation of a binary black hole begins by solving the Einstein constraint equations on the initial time slice.
For conformally flat, maximally sliced puncture data, this is a nonlinear elliptic boundary-value problem: the Hamiltonian constraint for the conformal factor in Lichnerowicz--York form~\cite{York1979,BowenYork1980,BrandtBruegmann1997,Cook2000}.
This problem must be solved once for each configuration.

Modern applications, however, rarely require a single solution.
Parameter surveys, waveform catalogs, eccentricity reduction, and mass and spin targeting all require the same elliptic problem to be solved repeatedly across the binary-black-hole parameter space.
The repeated solution of the constraint equations therefore becomes a recurring computational cost~\cite{Mendes2025}.
The computational challenge is therefore not solving a single elliptic problem, but efficiently solving a family of closely related elliptic problems across the physical parameter space.

Reduced-order modeling is a mature response to exactly this kind of repeated computation.
In gravitational-wave physics, however, it has been developed almost exclusively for \emph{waveforms}: reduced-basis template banks and surrogate models reproduce waveforms across para\-meter space at a tiny fraction of the cost~\cite{FieldGalley2011,Field2014,Blackman2015,Varma2019}.
The elliptic initial-data solve that precedes every numerical-relativity evolution has largely remained outside this development.
Reduced-basis emulation does exist for the neutron-star sector, for example through Tolman--Oppenheimer--Volkoff emulators over equation-of-state parameters~\cite{Reed2024}, but no comparable parametric solver exists for binary-black-hole initial data.

From the perspective of numerical analysis, this gap is somewhat surprising.
Parametric collocation, sparse interpolation, and reduced-basis methods for parameter-dependent elliptic partial differential equations are well-established~\cite{BabuskaNobileTempone2007,XiuHesthaven2005,NobileTemponeWebster2008,Haasdonk2017,HesthavenRozzaStamm2016}.
When the parameter-to-solution map is analytic, the interpolation error decays exponentially with the number of samples at a rate determined by the nearest singularity in the complex para\-meter plane through the Bernstein-ellipse estimate~\cite{Trefethen2019}.
These results suggest that binary-black-hole puncture initial data should also admit efficient parametric representations over suitable regions of parameter space.

Recent developments have addressed complementary aspects of this problem.
Mendes \emph{et al.}~\cite{Mendes2025} accelerate para\-meter targeting through a Broyden iteration that repeatedly re-solves the constraint equations.
Tommasini \emph{et al.}~\cite{Tommasini2026} use Gaussian-process regression trained on the SXS catalog to predict low-eccentricity orbital para\-meters, reducing the number of trial evolutions required for eccentricity reduction.
Zhou \emph{et al.}~\cite{Zhou2026} introduce a physics-informed neural-network solver for the puncture Hamiltonian constraint on a single physical configuration and identify parameterized initial-data solvers as an important direction for future work.
Analytic perturbative constructions~\cite{Ogurol2025} and neural representations of spacetime metrics~\cite{Cranganore2025} provide additional complementary approaches.
None of these methods, however, constructs a parametric solver that can certify the constraint satisfaction, and provide analytic derivatives of the resulting initial data with respect to the physical parameters.

This paper addresses this gap by introducing a parametric solver for binary-black-hole puncture initial data that is analytically differentiable with respect to the physical parameters and whose every returned datum satisfies the constraints to a checked tolerance.
Rather than replacing the elliptic solve with an approximation, the method constructs a sparse interpolant that provides a high-quality initial guess for the nonlinear solve, after which a small number of Newton iterations recover a fully converged solution satisfying the Einstein constraint equations.
We demonstrate the approach for both a four-dimensional aligned-spin model and an eight-dimensional general-spin quasi-circular model, and apply it to two representative tasks: targeting physical parameters and reducing orbital eccentricity.

The proposed solver is \emph{parametric} and \emph{analytically differentiable} with respect to the physical parameters, a combination that, to our knowledge, has not previously been demonstrated for binary-black-hole initial data.
The entire construction is implemented in JAX~\cite{Jax2018}, so the resulting parameter-to-solution map is differentiable with respect to the physical parameters, and these derivatives are obtained automatically through algorithmic differentiation rather than by finite differencing.
Neither property is bought at the cost of accuracy.
Every queried parameter is refined by Newton iterations of the genuine elliptic solver until the discrete, row-equilibrated constraint residual (Sec.~\ref{sec:solve:newton}) satisfies $\|R\|_\infty \le 10^{-10}$, a gate checked before the datum is returned; the corresponding continuum constraint violation is verified independently on an evolution grid (App.~\ref{sec:validation:constraints}).
The sparse interpolant therefore accelerates the computation but does not replace the underlying elliptic solver.

The remainder of this paper is organized as follows.
Section~\ref{sec:problem} reviews the puncture initial-data formulation and the resulting nonlinear elliptic constraint equation.
Section~\ref{sec:solve} describes the spectral elliptic solver that produces a single constraint-satisfying datum.
Section~\ref{sec:param} builds the parametric solver over it, combining sparse interpolation, Newton refinement, and a differentiable JAX implementation.
Section~\ref{sec:model} characterizes the aligned-spin and general-spin quasi-circular models, including their convergence and the analyticity limits of the para\-meter map.
Section~\ref{sec:applications} applies the method to certified para\-meter targeting and eccentricity reduction.
Finally, Section~\ref{sec:conclusions} discusses the limitations, implications, and conclusions of this work.
Validation of the initial data against TwoPunctures is collected in Appendix~\ref{sec:validation}.

\section{Puncture initial data}
\label{sec:problem}
We briefly review the puncture formulation of binary-black-hole initial data used
throughout this work. We adopt conformally flat, maximally sliced initial data
within York's conformal transverse--traceless
de\-com\-position~\cite{York1979,Cook2000}, rather than the conformal thin-sandwich
formulation~\cite{York1999}. Under these assumptions, the momentum constraint is
satisfied analytically, leaving a single nonlinear elliptic equation for the
conformal factor.

\subsection{Constraint and puncture decomposition}
\label{sec:problem:decomp}

In the conformal decomposition, the spatial metric and extrinsic curvature are
$\gamma_{ij} = \psi^4 f_{ij}$ and $K^{ij} = \psi^{-10}\Ahat^{ij}$, where
$f_{ij}$ is the flat conformal metric and $\Ahat^{ij}$ is the conformal
(Bowen--York) extrinsic curvature. We assume maximal slicing,
$K\equiv\gamma_{ij}K^{ij}=0$, equivalently $f_{ij}\Ahat^{ij}=0$, since
$K=\psi^{-6}f_{ij}\Ahat^{ij}$. If $\Ahat^{ij}$ is transverse with respect to the flat
metric, $\partial_i\Ahat^{ij}=0$, the momentum constraint is satisfied
analytically (Sec.~\ref{sec:problem:source}), while the Hamiltonian constraint
reduces to the non\-linear Lichnerowicz--York equation
\begin{equation}
  \Delta \psi \ = \ -\tfrac{1}{8}\, \psi^{-7}\, \Ahat_{ij}\Ahat^{ij},
  \label{eq:lichnerowicz}
\end{equation}
where $\Delta$ is the flat Laplacian and indices are raised and lowered with
$f_{ij}$. This is the only equation that must be solved numerically.

Equation~\eqref{eq:lichnerowicz} is defined on an unbounded domain and is
singular at each puncture because $\Ahat^2$ diverges there. Following Brandt and
Br\"ugmann~\cite{BrandtBruegmann1997}, we decompose the conformal factor as
$\psi=\psiBL+u$, where the $N$-puncture Brill--Lindquist background is
\begin{equation}
  \psiBL \ = \ 1 \ + \ \sum_{X=1}^{N} \frac{m_X}{2 r_X},
  \label{eq:psiBL}
\end{equation}
where $r_X \equiv |\vect{x}-\vect{x}_X|$ is the coordinate distance to
puncture $X$. The Brill--Lindquist term contains all $1/r_X$ singularities and
is harmonic away from the punctures. The regular correction therefore satisfies
\begin{equation}
  \Delta u \ = \ -\tfrac{1}{8}\,(\psiBL + u)^{-7}\, \Ahat_{ij}\Ahat^{ij},
  \label{eq:u_eq}
\end{equation}
where, for the binary, we take $N=2$ punctures of bare masses $m_A$ and $m_B$
located at $z=\pm b$. Solving Eq.~\eqref{eq:u_eq} for $u$ determines the
complete initial data.

\subsection{Bowen--York source term}
\label{sec:problem:source}
The momentum constraint is satisfied analytically by the Bowen--York extrinsic
curvature~\cite{BowenYork1980},
\begin{align}
  \Ahat^{ij} \ &= \ \sum_{X\in\{A,B\}} \bigl(\Ahat^{ij}_{P,X} + \Ahat^{ij}_{S,X}\bigr),
\end{align}
where the momentum and spin contributions are
\begin{align}
  \Ahat^{ij}_{P,X} \ &\equiv \ \frac{3}{2 r_X^2}\Bigl(
  P_X^i n_X^j + P_X^j n_X^i \nonumber \\
  & \hspace{15mm} - \bigl(f^{ij} - n_X^i n_X^j\bigr)(\vect{P}_X\!\cdot\!\vect{n}_X)\Bigr),
  \label{eq:A_P}
  \\
  \Ahat^{ij}_{S,X} \ &\equiv \ \frac{3}{r_X^3}\bigl(v_X^i n_X^j + v_X^j n_X^i\bigr),
  \label{eq:A_S}
\end{align}
where $\vect{P}_X$ and $\vect{S}^{X}$ are the linear momentum and spin of
puncture $X$, respectively,
$\vect{v}_X \equiv \vect{S}^{X}\times\vect{n}_X$, and
$\vect{n}_X \equiv (\vect{x}-\vect{x}_X)/r_X$ is the unit
radial vector centered on puncture $X$. The tensor $f^{ij}$ in Eq.~\eqref{eq:A_P} is the inverse flat
conformal metric; in the Cartesian coordinates used to construct the tensor,
$f^{ij}=\delta^{ij}$. Each Bowen--York contribution is transverse and
trace-free with respect to $f_{ij}$. Their sum therefore satisfies the momentum
constraint analytically for arbitrary momenta, spins, and separations.

The Hamiltonian constraint~\eqref{eq:u_eq} depends only on the scalar contraction
$\Ahat_{ij}\Ahat^{ij}$. Defining
$\Ahat_X \equiv \Ahat_{P,X}+\Ahat_{S,X}$ as the total tensor for
puncture $X$, the contraction decomposes into two self-contractions and a cross
term,
\begin{equation}
  \Ahat_{ij}\Ahat^{ij} \ = \ \Ahat_A\!:\!\Ahat_A \;+\; 2\,\Ahat_A\!:\!\Ahat_B \; + \; \Ahat_B\!:\!\Ahat_B .
  \label{eq:A2_full}
\end{equation}
Each self-contraction is
\begin{equation}
\begin{aligned}
  \Ahat_X\!:\!\Ahat_X \ &= 
  \ \frac{9}{2r_X^4}\bigl(P_X^2 \, + \, 2(\vect{P}_X\!\cdot\!\vect{n}_X)^2\bigr) \\
  & \hspace{5mm} \ + \ \frac{18}{r_X^6}\,|\vect{S}^{X}\!\times\!\vect{n}_X|^2 \\
  & \hspace{5mm} \ + \ \frac{18}{r_X^5}\,(\vect{P}_X\!\times\!\vect{S}^{X})\!\cdot\!\vect{n}_X ,
\end{aligned}
  \label{eq:A2_self}
\end{equation}
where the three contributions are momentum, spin, and momentum--spin,
respectively. The cross term is
\begin{equation}
\begin{aligned}
  2\,\Ahat_A\!:\!\Ahat_B
  \ &= \ 
  \frac{9}{2r_A^2 r_B^2}\,\Pi_{AB}
  \ + \ \frac{36}{r_A^3 r_B^3}\,\Sigma_{AB} \\
  & \hspace{5mm} + \ \frac{18}{r_A^2 r_B^3}\,\Xi_{AB}
  \ + \ \frac{18}{r_A^3 r_B^2}\,\Xi_{BA} ,
\end{aligned}
  \label{eq:A2_cross}
\end{equation}
where the momentum--momentum ($\Pi$), spin--spin ($\Sigma$), and
momentum--spin ($\Xi$) invariants are
\begin{align}
  \Pi_{AB} \ &\equiv \ 2(\vect{P}_A\!\cdot\!\vect{n}_B)(\vect{P}_B\!\cdot\!\vect{n}_A)
   - 3(\vect{P}_A\!\cdot\!\vect{n}_A)(\vect{P}_B\!\cdot\!\vect{n}_B) \nonumber\\
  &\hspace{5mm} + 2(\vect{n}_A\!\cdot\!\vect{n}_B)\bigl[ (\vect{P}_A\!\cdot\!\vect{P}_B) \nonumber \\
  &\hspace{25mm} + (\vect{P}_A\!\cdot\!\vect{n}_A)(\vect{P}_B\!\cdot\!\vect{n}_A) \nonumber\\
  &\hspace{25mm} + (\vect{P}_B\!\cdot\!\vect{n}_B)(\vect{P}_A\!\cdot\!\vect{n}_B)\bigr] \nonumber\\
  &\hspace{ 5mm} + (\vect{P}_A\!\cdot\!\vect{n}_A)(\vect{P}_B\!\cdot\!\vect{n}_B)(\vect{n}_A\!\cdot\!\vect{n}_B)^2 ,
  \label{eq:invPi}\\
  \Sigma_{AB} \ &\equiv \ (\vect{v}_A\!\cdot\!\vect{v}_B)(\vect{n}_A\!\cdot\!\vect{n}_B) + (\vect{v}_A\!\cdot\!\vect{n}_B)(\vect{v}_B\!\cdot\!\vect{n}_A) ,
  \label{eq:invSigma}\\
  \Xi_{AB} \ &\equiv \ (\vect{P}_A\!\cdot\!\vect{v}_B)(\vect{n}_A\!\cdot\!\vect{n}_B)
   + (\vect{P}_A\!\cdot\!\vect{n}_B)(\vect{n}_A\!\cdot\!\vect{v}_B) \nonumber\\
  & \hspace{5mm} + (\vect{P}_A\!\cdot\!\vect{n}_A)(\vect{n}_A\!\cdot\!\vect{n}_B)(\vect{n}_A\!\cdot\!\vect{v}_B) ,
  \label{eq:invXi}
\end{align}
and where $\Xi_{BA}$ follows from exchanging $A$ and $B$.

\subsection{Regularity of the puncture formulation}
\label{sec:problem:regularity}
The puncture decomposition isolates the singular part of the conformal factor in the analytic Brill--Lindquist background $\psiBL$.
Near puncture $X$, $\psiBL^{-7}\sim r_X^7$, whereas the momentum and spin self-contractions scale as $\Ahat_{P,X}^2\sim r_X^{-4}$ and $\Ahat_{S,X}^2\sim r_X^{-6}$, respectively.
Hence, their contributions to the effective source in Eq.~\eqref{eq:u_eq} vanish at least linearly as $r_X\to0$.
The cross terms are likewise regular and vanish at the puncture.
A source vanishing linearly leaves $u$ with a local $r_X^3$ term, which in Cartesian coordinates is only twice differentiable at the puncture.
The obstruction is one of the chart rather than of $u$ itself, and the coordinates introduced in Sec.~\ref{sec:solve:abt} are chosen to remove it.

\section{Constraint solver}
\label{sec:solve}
We solve Eq.~\eqref{eq:u_eq} for the regular correction $u$ by spectral collocation on a single spectral patch.
We use a single-patch pseudo-spectral method in the manner of Ansorg--Br\"ugmann--Tichy (TwoPunctures)~\cite{AnsorgBruegmannTichy2004}, with an equivalent prolate-spheroidal chart and our own collocation nodes and boundary rows.
This solver provides the reference solution on which the parametric method of Sec.~\ref{sec:param} is built and against which it is validated.

\subsection{The prolate-spheroidal single patch}
\label{sec:solve:abt}

Following the single-domain construction of Ansorg--Br\"ugmann--Tichy (ABT)~\cite{AnsorgBruegmannTichy2004}, we map the two-puncture domain onto a single compact spectral patch.
Starting from cylindrical co\-ordinates $(\rho,z,\phi)$, we use an equivalent compactified prolate-spheroidal parameterization with its foci at the punctures $z=\pm b$, whose coordinate separation is $2b$.
With $A\in[0,1]$ and $B\in[-1,1]$, the transformation to cy\-lindrical coordinates is
\begin{equation}
  \frac{\rho}{b} \ = \ \frac{2A\sqrt{1-B^2}}{1-A^2}, \qquad
  \frac{z}{b}    \ = \ \frac{(1+A^2)\,B}{1-A^2}.
  \label{eq:abt_map}
\end{equation}
The coordinate edges have a clean meaning.
The edge $A=1$ represents spatial infinity, the edge $A=0$ maps to the inner segment $|z|\le b$ of the azimuthal axis, and $B=\pm1$ map to its two outer segments $|z|\ge b$.
Their intersections $(A=0,B=\pm1)$ are the punctures.
These are degeneracies of the cylindrical chart rather than physical inner boundaries; the construction introduces no excision surface.

We place $N_A{+}1$ Chebyshev--Gauss--Lobatto nodes in $A$, including both edges $A{=}1$ and $A{=}0$; $N_B$ interior Gauss--Legendre nodes in $B$, which avoid the polar edges $B=\pm1$; and $N_\phi$ equispaced nodes in $\phi$.
The $B$ nodes are retained in full, positive and negative alike, so no $z$-reflection symmetry is imposed.
Because the $B$ edges carry no node, the puncture corners are not collocation points.
The regular correction $u$ is represented nodally, the unknown being the tensor of collocation values $U_{ijk}\equiv u(A_i,B_j,\phi_k)$.

The solver stores this nodal tensor, but the approximation space it spans is most transparent in modal form,
\begin{equation}
  u(A,B,\phi) \ = \sum_{|m|\le N_\phi/2} \ \sum_{n=0}^{N_A}\sum_{l=0}^{N_B-1} c^{m}_{nl} \ \phi_{nl}^{m}(A,B,\phi) \label{eq:expansion}
\end{equation}
with basis functions,
\begin{equation}
  \phi_{nl}^{m}(A,B,\phi) \, \equiv \, T_n(2A-1) \, P_l(B) (1-B^2)^{|m|/2} \, e^{im\phi} .
  \label{eq:basis}
\end{equation}
The basis consists of a Fourier expansion in the azimuthal angle $\phi$ (with
$c^{-m}_{nl}=(c^{m}_{nl})^{*}$ since $u$ is real, and, for even $N_\phi$, with the
Nyquist term $|m|=N_\phi/2$ counted once, since on $N_\phi$ equispaced nodes it
retains only its cosine part; see also Fig.~\ref{fig:tpspec}), Chebyshev polynomials $T_n$ in the compactified radial coordinate
$A$, and Legendre polynomials $P_l$ in the angle $B$.
The implementation stores the real-FFT half-spectrum $m=0,\dots,N_\phi/2$, the negative modes following from the reality condition.
The coefficients $c_{nl}^m$ are not formed explicitly.
The azimuthal FFT maps the $N_\phi$ node values to the modes $m$, and within each meridian block the $A$- and $B$-derivatives act on the nodal values through the Chebyshev and Legendre differentiation matrices.

Two regularity requirements shape this basis.
Regularity on the azimuthal axis requires the $m$th Fourier mode to vanish as $\rho^{|m|}$.
At fixed $A$, $\rho\propto\sqrt{1-B^2}$ near $B=\pm1$, so we write $u_m=w_m\,(u_m/w_m)$ with $w_m(B)\equiv(1-B^2)^{|m|/2}$ and $u_m/w_m$ smooth in $B$; this is the associated-Legendre prefactor of Eq.~\eqref{eq:basis}.
For odd $m$ the square root is coordinate behavior at the axis rather than a physical singularity, and a bare polynomial resolves it only algebraically.
We therefore differentiate $w_m$ analytically and apply the polynomial differentiation matrices only to the smooth factor, restoring spectral convergence.

The chart also removes the puncture obstruction of Sec.~\ref{sec:problem:regularity}.
Inverting Eq.~\eqref{eq:abt_map} gives the distance to each puncture,
\begin{equation}
  \frac{r_{\pm}}{b} \ = \ \frac{A^2(1\pm B) + (1\mp B)}{1-A^2},
  \label{eq:puncture_distance}
\end{equation}
with $r_{+}$ the distance to the puncture at $z=+b$ and $r_{-}$ that to the puncture at $z=-b$.
Each is rational in the chart variables and vanishes only at its own corner, so the odd powers of $r_X$ generated there, including the $r_X^3$ term responsible for the limited Cartesian differentiability, are analytic at that corner.
The limited Cartesian differentiability at a puncture therefore does not limit regularity in $(A,B)$ there, and together with $w_m$ the representation also enforces regularity on the azimuthal axis.
It does not make $u$ smooth on the whole patch: for nonvanishing puncture momenta the solution carries logarithmic terms at the $A=1$ edge, entering at order $(1-A)^5\log(1-A)$ when the total linear momentum vanishes~\cite{AnsorgBruegmannTichy2004}.
Regularity at the punctures is what the chart buys; the edge at infinity sets the rate at which the expansion of Eq.~\eqref{eq:expansion} converges.

Because the ABT coordinates are orthogonal, the flat-space Laplacian contains no mixed $A$--$B$ derivatives,
\begin{equation}
\begin{aligned}
   \Delta u \ &= \ \frac{(1-A^2)^2}{b^2\,\mathcal{D}} \bigg[
    \frac{(1-A^2)^2}{4A}\,\partial_A\!\big(A\,\partial_A u\big) \\
    &\hspace{23mm} \, + \, \partial_B\!\big((1-B^2)\,\partial_B u\big)\bigg]
     + \frac{\partial_{\phi}^2 u}{\rho^{2}},
\end{aligned}
  \label{eq:abt_laplacian}
\end{equation}
with $\mathcal{D} \equiv (1+A^2)^2 - B^2(1-A^2)^2$ and $\rho$ from
Eq.~\eqref{eq:abt_map}.
True Kronecker separability requires the further row scaling of Sec.~\ref{sec:solve:newton}.

In a Fourier basis in $\phi$ the azimuthal derivative dia\-gonalizes, $\partial_\phi^2\!\rightarrow\!-m^2$, so the discretized flat Laplacian is block diagonal in $m$, with one two-dimensional $(A,B)$ block per retained mode.
The full three-dimensional dense matrix is therefore never assembled.
The nonlinear source is evaluated on the azimuthal collocation grid and generally couples these blocks; Sec.~\ref{sec:solve:newton} treats that coupling in the full Newton Jacobian.

Boundary and regularity conditions are imposed by row replacement.
The only asymptotic boundary condition is $u=0$ at $A=1$.
The rows at $A=0$ enforce regularity on the inner axis: the axisymmetric mode satisfies $\partial_A u_0 = 0$, while every regular $m\neq0$ mode vanishes there $u_m=0$.
No separate puncture or excision condition is imposed.
Likewise no rows are imposed at $B=\pm1$: the Gauss--Legendre nodes exclude those edges, and the factor $w_m$ supplies their regular behavior.

Let $M_0^{(m)}$ denote the resulting row-replaced Laplacian block for mode $m$.
Its rows span a huge
dynamic range: the $(1-A^2)^4$ factor is tiny near infinity and the $1/A$ factor
large near the inner axis, giving a raw condition number $\sim10^{13}$--$10^{15}$.
Left-scaling each row by the reciprocal of its largest-magnitude entry, an exact
rescaling of the equations that leaves the discrete solution unchanged, collapses the
condition number to $\sim10^4$ and permits accurate row-equilibrated linear
solves.

A solved field must also be evaluated \emph{off} the collocation grid.
For an off-grid query we first invert Eq.~\eqref{eq:abt_map} to obtain $(A,B)$.
The nodal values are transformed to azimuthal modes, each mode is divided by its factor $w_m$, and the smooth factor is interpolated in $A$ and $B$ by the numerically stable barycentric-Lagrange formula~\cite{Trefethen2019}; $w_m$ is then restored analytically at the query point and the Fourier series is summed there.
This is not an approximation to the truncated spectral representation.
The barycentric interpolant is the unique polynomial of the retained degree through the nodal values of the smooth factor, and the trigonometric sum is the corresponding band-limited periodic interpolant, so the evaluated field is the expansion of Eqs.~\eqref{eq:expansion} and~\eqref{eq:basis} in coefficient-free form.
Dividing out $w_m$ is what makes that identity hold for every mode: the physical mode $u_m$ is not a polynomial in $B$ (for odd $m$ not even smooth at $B=\pm1$), so interpolating the nodal values directly would reproduce the expansion only at $m=0$.

\subsection{Newton iteration with an analytic Jacobian}
\label{sec:solve:newton}

Collocation on the grid of Sec.~\ref{sec:solve:abt} reduces
Eq.~\eqref{eq:u_eq} to a finite nonlinear system $R(U)=0$, where
$U_{ijk}=u(A_i,B_j,\phi_k)$. We solve this system with an outer Newton
iteration. Each Newton equation is solved by the generalized minimal residual
(GMRES) method using a matrix-free Jacobian action and a mode-wise Laplacian
preconditioner. The full coupled three-dimensional Jacobian is never assembled,
although the single-mode $N_\phi=1$ path constructs its block Jacobian as the
preconditioner.

\paragraph{Newton linearization.}
Starting from an initial iterate $U^{(0)}$, each Newton step solves the
linearized system
\begin{equation}
  J\bigl(U^{(k)}\bigr)\,\delta U^{(k)} \ = \ -\,R\bigl(U^{(k)}\bigr),
  \label{eq:newton}
\end{equation}
for the increment $\delta U^{(k)}$ and updates
$U^{(k+1)} = U^{(k)} + \delta U^{(k)}$.
The residual and its Jacobian $J = \partial{R} / \partial U$ are
\begin{align}
  R(U) \ &\equiv \ M_0\,U \ + \ \tfrac{1}{8}\,(\psiBL+U)^{-7}\,\Ahat^2 \big|_{\rm int},
  \label{eq:residual}
  \\
  J(U) \ &\equiv \ M_0 \ - \ \tfrac{7}{8}\,\mathrm{diag}\!\big[(\psiBL+U)^{-8}\,\Ahat^2\big]_{\rm int},
  \label{eq:jacobian}
\end{align}
where $M_0$ is the discretized Laplacian of Eq.~\eqref{eq:abt_laplacian},
including the row-replaced boundary conditions, and the subscript ``int''
indicates that the nonlinear source enters only on the interior rows. The
derivative of the nonlinear source is multiplication by the local coefficient
\begin{equation}
D_{\rm nl}(A,B,\phi)
\ \equiv \
-\tfrac78(\psiBL+U)^{-8}\Ahat^2,
\end{equation}
which is diagonal in the nodal basis; the boundary rows of $J$ remain those of
$M_0$. We evaluate the nonlinear term pseudo-spectrally on the $\phi$
collocation grid.

\paragraph{Matrix-free Fourier solve.}
We solve the linear system at each Newton step with GMRES. The two terms in the Jacobian
Eq.~\eqref{eq:jacobian} are simplest in different representations. In Fourier
space, $M_0$ is block diagonal in $m$, since
$\partial_\phi^2\!\rightarrow\!-m^2$, and each block is a two-dimensional
operator on $(A,B)$. By contrast, $D_{\rm nl}$ generally depends on $\phi$.
Pointwise multiplication by this coefficient corresponds to convolution in
Fourier space, coupling the $m$-blocks, so $J$ is generally not block diagonal.

GMRES requires only the action $v\mapsto Jv$, evaluated term by term in its
natural representation,
\begin{equation}
  \widehat{(Jv)}^{(m)}
  \ = \
  \widehat{(M_0v)}^{(m)}
  \ + \ \Big\{\mathcal{F}_\phi\big[
      D_{\rm nl}\,\mathcal{F}_\phi^{-1}(\hat v)
    \big]\Big\}^{(m)}_{\rm int},
  \label{eq:jacobian_action}
\end{equation}
with $\hat v=\mathcal{F}_\phi(v)$. The stored factored block evaluates the first
term by acting on $\hat v^{(m)}/w_m$, where $w_m(B)=(1-B^2)^{|m|/2}$ is the
associated-Legendre factor of Eq.~\eqref{eq:basis}. The second term is
transformed to the collocation grid, multiplied pointwise by $D_{\rm nl}$, and
transformed back. Equation~\eqref{eq:jacobian_action} applies the exact discrete
Jacobian of Eq.~\eqref{eq:jacobian}; ``matrix-free'' describes how $J$ is applied
and introduces no approximation.

\paragraph{Separable preconditioner.}
Matrix-free application reduces storage but does not by itself ensure rapid GMRES
convergence. In the default $N_\phi>1$ path, we precondition with the inverse of
$M_0$ alone, omitting the entire multiplier $D_{\rm nl}$, both its
mode-preserving and its mode-coupling parts. Because GMRES still uses the full
Jacobian action, this approximation affects the convergence rate but not the
Newton system or its converged solution.

Each Fourier block $M_0^{(m)}$ of the linear operator becomes separable after a
simple row scaling. Dividing its interior rows by the common prefactor
$(1-A^2)^2/(b^2\mathcal{D})$ in Eq.~\eqref{eq:abt_laplacian} removes the only
coefficient that depends jointly on $A$ and $B$. The centrifugal term separates in the same way: with $\rho$ from
Eq.~\eqref{eq:abt_map},
\begin{equation}
  \frac{m^2}{\rho^2}\bigg/\frac{(1-A^2)^2}{b^2\,\mathcal{D}}
 \ = \ m^2\left[\,\frac{1}{1-B^2} \ + \ \frac{(1-A^2)^2}{4A^2}\,\right],
  \label{eq:centrifugal_split}
\end{equation}
a function of $B$ alone plus a function of $A$ alone. Together with the
associated-Legendre factor, the scaled block factors as
$(I\otimes W)\big(L_A^{(m)}\otimes I+I\otimes L_B^{(m)}\big)$, where
$W=\mathrm{diag}(w_m)$ and $L_A^{(m)}$ and $L_B^{(m)}$ are one-dimensional
operators in $A$ and $B$: a diagonal factor times a true Kronecker sum. Fast
diagonalization then
replaces a full two-dimensional factorization with two one-dimensional
eigendecompositions per mode. Appendix~\ref{sec:separable} gives the algebra and
boundary treatment.

Although $M_0$ contains the coordinate parameter $b$, this dependence enters only
through a diagonal left scaling that row equilibration cancels. The
one-dimensional factors therefore depend only on the grid and can be constructed
once and reused for all physical configurations and Newton iterates. Per mode,
storage falls from $O\big((N_AN_B)^2\big)$ to $O(N_A^2+N_B^2)$. Their parameter
independence also means that differentiation in Sec.~\ref{sec:param:diff} need
not pass through the eigendecompositions.

For $N_\phi=1$, we instead retain the nonlinear diagonal in the block
preconditioner. There is then no azimuthal coupling, so the block is the full
Jacobian and preconditioned GMRES converges already in a single iteration.

\paragraph{Convergence and certification.}
Residual monitoring requires care because the rows of the prolate operator span
the wide dynamic range described in Sec.~\ref{sec:solve:abt}. The raw nodal norm
$\|R\|_\infty$ is dominated by poorly scaled rows near the inner axis; it
saturates well above machine precision and can even increase with resolution.
This behavior reflects row scaling rather than poorer constraint satisfaction. We
therefore measure each discrete equation against the magnitude of its own
operator row,
\begin{align}
  s^{(m)}_i
  \ &\equiv \
  \max_j \left| \left(M_0^{(m)}\right)_{ij}\right| ,
  \label{eq:row_scale} \\
  \|R\|_{\infty,\rm eq}
  \ &\equiv \
  \max_m \max_i
     \bigl| \hat R^{(m)}_{i} \big/ s^{(m)}_{i} \bigr| .
  \label{eq:equil_residual}
\end{align}
Because row equilibration is only a left scaling, it does not change the root of
the discrete system. We therefore monitor Eq.~\eqref{eq:equil_residual} and
abbreviate it as $\|R\|^{(v)}_\infty$ throughout, the superscript naming the
row-equilibrated normalization of Eq.~\eqref{eq:row_scale} and distinguishing it
from the raw nodal norm of the preceding paragraph. Every residual and every
tolerance quoted in this paper is in that normalization.\footnote{The row scale of
Eq.~\eqref{eq:row_scale} is not the only defensible choice, and the distinction
matters quantitatively. A normalization that divides each equation by the
magnitude of its own \emph{column} rather than by the maximum over its row --
denoted $\|R\|^{(u)}_\infty$ -- removes a boundary factor that grows without bound
with the number of azimuthal modes, and which causes
Eq.~\eqref{eq:equil_residual} to reject field-converged solutions at high
$N_\phi$ while accepting coarser ones. The implementation accompanying this work
has since adopted that normalization as its default monitor and certifies against
$\|R\|^{(u)}_\infty\le10^{-11}$. \emph{This is not a tightening of the gate used
here; it is substantially looser.} The two normalizations are not related by a
fixed factor: their ratio is bounded by the grid-dependent quantity
$(1-\max_j B_j^2)^{-m_{\max}/2}$, which equals $3.3\times10^{4}$ at the
production grid $44\times32\times8$ of this paper and grows without bound with
$N_\phi$, while the ratio attained on an individual converged solve is smaller
and configuration-dependent. Either way the newer gate admits residuals orders of
magnitude larger when expressed in $\|R\|^{(v)}_\infty$, so no number in this
paper should be compared with one quoted against
$\|R\|^{(u)}_\infty\le10^{-11}$ without that conversion. Every value reported
here is the measured $\|R\|^{(v)}_\infty$ and is unchanged by that later
choice.}

The nonlinear iteration drives the residual to an internal target one decade
below the requested tolerance, and terminates when that target is reached, when
the best residual fails to halve on two consecutive iterations, or when the
budget of Newton steps is exhausted, returning the lowest-residual iterate
encountered. Termination alone does not constitute
certification: the interface of Sec.~\ref{sec:polish} returns a datum only if
Eq.~\eqref{eq:equil_residual} satisfies the prescribed tolerance; otherwise the
attempt remains uncertified. This equilibrated residual is an internal diagnostic
of the discrete solver. Appendix~\ref{sec:validation:constraints} separately
evaluates the physical Hamiltonian-constraint violation on an evolution grid
using an independent numerical-relativity code.

This elliptic solver refines any sufficiently accurate guess to the
prescribed residual tolerance. The next section describes how an improved guess can be
constructed.

\section{Parametric solver}
\label{sec:param}
So far we have described how to obtain one constraint-satisfying datum. The
contribution of this paper is to obtain the datum \emph{as a function of} the
physical para\-meters, without a cold elliptic solve at query time.

\subsection{The parametric problem}
\label{sec:param:setup}

We consider the parameter vector $\theta \equiv (q, b, \boldsymbol{\chi}^{A}, \boldsymbol{\chi}^{B})$ at fixed total
bare mass $M \equiv m_A + m_B$ (set to $1$), with the spins carried as the
Kerr-like \emph{dimensionless} spins $\boldsymbol{\chi}^{X} \equiv \vect{S}^{X}/m_X^2$, each a Cartesian
3-vector referred to the bare puncture mass $m_X$.
The per-puncture momenta are not interpolation axes but are determined by the
quasi-circularity condition of Sec.~\ref{sec:model:qc}.
The map to a slice is algebraic: $m_A = Mq/(1+q)$, $m_B = M/(1+q)$, and the
Bowen--York spin that enters the analytic source is $\vect{S}^{X} = \boldsymbol{\chi}^{X}\, m_X^2$. The
interpolation axes are $q$, $b$, and the individual spin \emph{components}
$\chi^{X}_{i}$, i.e.\ up to eight scalar axes for the full two-spin family.

Sampling the spins in $\chi^{X}_{i}$ keeps the parameter domain rectangular:
because $\boldsymbol{\chi}^{X}$ is normalized to its own puncture mass, its range
does not depend on the component masses, so the spin axes are fixed independent of
$q$ and symmetric about zero. The domain is thus a hypercube in the spin
\emph{components}.
Each interpolation axis is finally mapped affinely onto $[-1,1]$, the natural domain
of the Chebyshev basis.

\subsection{Value (Lagrange) collocation}
\label{sec:param:interp}

Every solution is stored on the same reference nodes $(A_i,B_j,\phi_k)$, so a
fixed nodal index denotes the same computational coordinates at every parameter
point. It does not denote the same Cartesian point when $b$ varies: the
reference nodes are fixed, but $\rho=b\,f(A,B)$ and $z=b\,g(A,B)$ by
Eq.~\eqref{eq:abt_map}. Parameter interpolation therefore acts on the fields
pulled back to this fixed reference patch, which is what makes tensor-product
interpolation over the parameters well-defined: the interpolant acts directly on
corresponding nodal values, producing an interpolated nodal tensor on the same
spectral grid. The resulting
parametric model inherits the smoothness of the underlying family of solutions.
Physical evaluation is a separate, subsequent step, applying the inverse map at
the queried value of $b$ and the coefficient-free spectral interpolation
described in Sec.~\ref{sec:solve:abt}.
The parametric model is then,
\begin{equation}
  \tilde{U}(\theta) \ = \
  \sum_{i_1=0}^{Q_1}\!\cdots\!\sum_{i_d=0}^{Q_d} \
  U_{i_1\cdots i_d} \,
  \prod_{k=1}^{d}
  L_{i_k}(\theta_k),
  \label{eq:surrogate}
\end{equation}
where $U_{i_1\cdots i_d}$ is the nodal tensor of the high-fidelity solve stored at
the parameter node
$(\theta_{1,i_1},\dots,\theta_{d,i_d})$ and $L_{i_k}$ is the
Lagrange cardinal function associated with the $k$-th parameter.
Equation~\eqref{eq:surrogate} defines the method: initial data at a new
parameter value are obtained as a linear combination of
the stored high-fidelity solutions.

We evaluate the cardinal functions in the numerically stable barycentric
form~\cite{Trefethen2019},
\begin{equation}
  L_{i_k}(\theta_{k})
  \ \equiv \
  \frac{\displaystyle \lambda_{i_{k}}/\big(\theta_k \, - \, \theta_{k,i_{k}}\big)}
       {\displaystyle \sum_{j_{k}=0}^{Q_k}\lambda_{j_{k}}/\big(\theta_k \, - \, \theta_{k,j_{k}}\big)},
  \label{eq:barycentric}
\end{equation}
where $\lambda_{i_k}=(-1)^{i_k}\delta_{i_k}$, with
$\delta_{i_k}=1/2$ at the endpoints and unity otherwise, so that
$L_{i_k}(\theta_{k,j_k})=\delta_{i_kj_k}$.
The nodes $\theta_{k,i_k}$ are the affine images of the
Chebyshev--Gauss--Lobatto points on
$[\theta_k^{\min},\theta_k^{\max}]$.

The tensor-product interpolant is evaluated by contracting one parameter axis at a time, requiring $O(\prod_k(Q_k+1))$ operations per query.
More important than the per-query cost is that the interpolant provides an
accurate initial guess throughout parameter space. Consequently, the certified
polish of Sec.~\ref{sec:polish} reaches tolerance in a median of two to three Newton
iterations, against four to five from a cold start.
Because the parameter-to-solution map is analytic, the
interpolation error decays exponentially in the number of parameter nodes, at a
rate fixed by the nearest singularity in the complex parameter plane.
This exponential convergence is demonstrated in
Sec.~\ref{sec:model:walls}.

\subsection{Certified refinement}
\label{sec:polish}
The parametric model of Eq.~\eqref{eq:surrogate} supplies only a
\emph{guess}. We certify it with a few Newton steps of the genuine elliptic
solver of Sec.~\ref{sec:solve:newton}. Taking the barycentric prediction as the
initial iterate, $U^{(0)} = \mathcal{I}[u](\theta)$, we then apply the Newton
iteration of Eq.~\eqref{eq:newton}, with the residual $R$ and analytic Jacobian $J$ of
Eqs.~(\ref{eq:residual}, \ref{eq:jacobian}). Because the guess is already close,
a few steps drive the constraint residual to $\|R\|^{(v)}_\infty \le 10^{-10}$,
\emph{independent} of the interpolation error, and that residual is checked before
the datum is returned, so a query cannot be silently wrong.
Section~\ref{sec:model:joint} makes this quantitative for the
shipped models.

Two quite different quantities measure how good a guess $U^{(0)}$ is: its
\emph{field} error $\|U^{(0)}-U_{\rm true}\|_2/\|U_{\rm true}\|_2$, the ordinary
distance between fields, which the interpolation accuracy controls; and the
\emph{constraint residual} $\|R(U^{(0)})\|_\infty$, which is what the Newton
iteration~\eqref{eq:newton} actually reduces. These are nearly \emph{decoupled}.
The residual operator differentiates the field (it contains the Laplacian, at
scale $\propto b^{-2}$), so it amplifies precisely the small-amplitude,
high-derivative content that the field norm barely registers, i.e.\ a guess can be
excellent in field-norm yet violate the constraint by orders of magnitude more.
Concretely, projecting the \emph{exact} solution onto a smooth low-dimensional
subspace leaves a field error $\sim\!10^{-4}$ but a residual
$\|R\|^{(v)}_\infty\sim\mathcal{O}(1)$, and it certifies in no fewer Newton steps than
the crude barycentric guess. The step count is therefore set by
$\|R(U^{(0)})\|^{(v)}_\infty$, not by field accuracy, with two consequences. First,
refining the interpolant's field accuracy is a weak lever on the online cost, so
the certified polish, which drives $\|R\|_{\infty}\to0$ regardless of the guess, is the
correct instrument. Second, one may discard the high-derivative content of the
fields with no penalty, since the polish reconstructs it; this enables the reduced-basis re-encoding of Sec.~\ref{sec:param:pod}.

\subsection{Differentiability}
\label{sec:param:diff}

The parametric solver is differentiable with respect to the physical parameters:
for any queried $\theta$ it exposes the sensitivities $\partial U/\partial\theta$
of the emitted field and, through any observable $F$ constructed from it, the
sensitivity $\partial F/\partial\theta$. That an interpolant is differentiable is
unremarkable; the content of the claim is that these are derivatives of the
\emph{certified} initial data. The challenge is not differentiating the
interpolant, but preserving differentiability through the projection onto
the constraint manifold. Implemented in JAX~\cite{Jax2018} as a
branchless barycentric map, the parametric model admits forward-mode automatic
differentiation, which returns $\partial U/\partial\theta$ and
$\partial F/\partial\theta$ directly, without numerical approximation.

At each node the tangent of the certified solution is available directly.
Differentiating the vanishing residual identity $R(U(\theta),\theta)\equiv0$ once gives the
first-order forward-sensitivity equation
\begin{equation}
  J\,\frac{\partial U}{\partial\theta_k} \ = \ -\,\frac{\partial R}{\partial\theta_k},
  \label{eq:tangent}
\end{equation}
with the same Jacobian $J=\partial R/\partial U$ (Eq.~\ref{eq:jacobian}) that the
certified solve uses. The desired parameter tangent, $\partial U / \partial{\theta_{k}}$, therefore reuses that matrix-free operator and its
block-diagonal preconditioner (Sec.~\ref{sec:solve:newton}), costing one
preconditioned GMRES solve and no further nonlinear solve.
What each axis needs is its own right-hand side, and on every axis of the
box that derivative is closed-form. On the interior rows
$\partial R/\partial\theta_k =
\tfrac{1}{8}\bigl[(\psiBL+u)^{-7}\,\partial\Ahat^2/\partial\theta_k
-7(\psiBL+u)^{-8}\,(\partial\psiBL/\partial\theta_k)\,\Ahat^2\bigr]$,
and it vanishes on the row-replaced boundaries. Because each Bowen--York tensor is
linear in its own $\vect{P}_X$ or $\vect{S}^{X}$, every $\partial\Ahat^2/\partial\theta_k$
is that same tensor re-evaluated with the derivative vector rather than a finite
difference of the source. The spins enter only through $\Ahat^2$, via
$S^{X} = \chi^{X} m_X^2$; the mass ratio through the bare-mass map,
$\partial\psiBL/\partial q = \sum_X (\mathrm{d}m_X/\mathrm{d}q)/2r_X$; and the separation
through the fixed-node scale laws $\partial\psiBL/\partial b = -(\psiBL-1)/b$ and
$\partial\Ahat^{ij}/\partial b = -(2/b)\Ahat^{ij}_{P}-(3/b)\Ahat^{ij}_{S}$, together with
the geometry term $-(2/b)\Delta u$ carried by the $b^{-2}$ scaling of the prolate
Laplacian, Eq.~\eqref{eq:abt_laplacian}. An orbiting family adds one term on $b$, $q$
and the aligned spins, since those axes also move the momenta through the
quasi-circularity condition of Sec.~\ref{sec:model:qc}: differentiating that closed-form
prescription gives $\mathrm{d}\vect{P}_X/\mathrm{d}\theta_k$, whose momentum tensor
enters $\partial\Ahat^2/\partial\theta_k$ alongside the terms above. Nothing in the
tangent is approximated.

The gradient the model exposes is therefore tied to the certified solve rather than
to the interpolant alone. On the axes enhanced in Sec.~\ref{sec:param:hermite} the
tangent of Eq.~\eqref{eq:tangent} is stored at every node, so the gradient exposed at
a node is that certified sensitivity identically. Away from the nodes, and on the axes
carrying plain Lagrange cardinals, the exposed gradient carries the interpolation error
of the derivative, which converges exponentially in the parameter order as the field
interpolation does. Differentiating an interpolant is not the same operation as
interpolating a derivative: this error is the derivative of the field interpolation
error, and is not bounded by it.

Certification and differentiability therefore coexist rather than compete: every
returned datum satisfies the constraints to tolerance, and an analytic parameter gradient is available at once, anchored at the nodes
to the certified sensitivity of Eq.~\eqref{eq:tangent}. This is what separates these
sensitivities from those of a bare interpolant or a neural-network surrogate, and
what makes the gradient-based applications of Sec.~\ref{sec:applications} possible.

\subsection{Gradient-enhanced (Hermite) collocation}
\label{sec:param:hermite}
Every certified solution carries verified parameter tangents, produced during the
offline build for the cost
of a single linear solve against the node's Jacobian
(Sec.~\ref{sec:param:diff}). This derivative information has already been
paid for, and to interpolate only the stored field values $U_i$ would discard it. Gradient-enhanced (Hermite) collocation
reincorporates it into the parametric model, matching values \emph{and}
parameter-derivatives at each node.

The gain is worth harvesting only where interpolation converges slowly. The fast
parameter directions already converge spectrally, so matching tangents helps most in
the slow directions near the Bernstein walls quantified in
Sec.~\ref{sec:model:walls}. We therefore enhance only a subset $\mathcal{E}$ of the
axes, and leave every other axis with its plain
Lagrange cardinal $L_{i_k}$ of Eq.~\eqref{eq:barycentric}. Along an enhanced axis
$k\in\mathcal{E}$ that cardinal is replaced by the Hermite pair
$(H_{i_k},\hat H_{i_k})$, and when two axes are enhanced together consistency
requires matching their mixed cross-derivative as well, giving the genuine
tensor-product Hermite interpolant on the enhanced subspace. For the two enhanced
axes used throughout this work, let $i_A$ and $i_B$ denote the two indices $i_k$
with $k\in\mathcal{E}$, write $U\equiv U_{i_1\cdots i_d}$ for the
certified solve stored at the node, and suppress the cardinal arguments,
$H_{i_A}\equiv H_{i_A}(\chi^{A}_{y})$ and $\hat H_{i_B}\equiv\hat H_{i_B}(\chi^{B}_{y})$.
The parametric model of Eq.~\eqref{eq:surrogate} then generalizes to
\begin{widetext}
\begin{equation}
  \tilde U(\theta) \ = \ \sum_{i_1=0}^{Q_1}\!\cdots\!\sum_{i_d=0}^{Q_d}
  \Bigg(\prod_{k\notin\mathcal{E}}\! L_{i_k}(\theta_k)\Bigg)
  \Big[\,
      H_{i_A}H_{i_B}\ U
    \, + \, \hat H_{i_A}H_{i_B}\ \partial_{\chi^{A}_{y}} U
    \, + \, H_{i_A}\hat H_{i_B}\ \partial_{\chi^{B}_{y}} U
    \, + \, \hat H_{i_A}\hat H_{i_B}\ \partial_{\chi^{A}_{y}}\partial_{\chi^{B}_{y}} U
  \,\Big],
  \label{eq:hermite}
\end{equation}
\end{widetext}
whose four terms are the value, the two certified first tangents, and the single
bilinear cross tangent stored at the node (Sec.~\ref{sec:param:diff}). Each
non-enhanced axis carries only its Lagrange cardinal, so restoring $L$ on the two
enhanced axes and dropping the tangents recovers
Eq.~\eqref{eq:surrogate}. The osculatory
Hermite cardinals are built from the same barycentric Lagrange cardinal $L_i$ of
Eq.~\eqref{eq:barycentric},
\begin{align}
  H_{i_k}(\theta) \ &\equiv \ \big[\,1 - 2\,L_{i_k}'(\theta_{i_k})\,(\theta-\theta_{i_k})\,\big]\,L_{i_k}(\theta)^2,
  \label{eq:hermite_cardinal_1} \\
  \hat H_{i_k}(\theta) \ &\equiv \ (\theta-\theta_{i_k})\,L_{i_k}(\theta)^2,
  \label{eq:hermite_cardinal_2}
\end{align}
with $L_i'(\theta_i)=\sum_{j\neq i}(\theta_i-\theta_j)^{-1}$ the diagonal of the
barycentric differentiation matrix. They satisfy
$H_{i_k}(\theta_{j_k})=\delta_{i_k j_k}$, $H_{i_k}'(\theta_{j_k})=0$,
$\hat H_{i_k}(\theta_{j_k})=0$, and $\hat H_{i_k}'(\theta_{j_k})=\delta_{i_k j_k}$, so
the interpolant reproduces the certified value \emph{and} tangents at every node by
construction, and the exposed gradient at the nodes equals the certified-solve
sensitivity exactly rather than the derivative of a value interpolant. Because the
parameter-to-solution map is analytic, matching the tangent raises the per-node
convergence order, so the same warm-start quality is reached with fewer expensive
elliptic solves.

For both shipped models the enhanced set is the pair of aligned spin components,
$\mathcal{E}=\{\chi^{A}_{y},\chi^{B}_{y}\}$, identified as the slowest axes in
Sec.~\ref{sec:model:peraxis}. The remaining axes carry plain Lagrange cardinals:
$b$ and $q$ in the four-dimensional aligned-spin model
$\theta=(b,q,\chi^{A}_{y},\chi^{B}_{y})$, and those two together with the four
in-plane components $\chi^{A}_{x},\chi^{A}_{z},\chi^{B}_{x},\chi^{B}_{z}$ in the
eight-dimensional general-spin model
$\theta=(b,q,\boldsymbol{\chi}^{A},\boldsymbol{\chi}^{B})$. The bracket of
Eq.~\eqref{eq:hermite} is therefore identical in the two models; only the number of
Lagrange factors differs.

Restricting enhancement to $\mathcal{E}$ also keeps the stored-derivative count
small: a full tensor-product Hermite over \emph{all} $d$ axes would carry mixed
partials up to order $d$ ($2^d$ fields per node), whereas Eq.~\eqref{eq:hermite}
carries only the $2^{|\mathcal{E}|}=4$ derivatives internal to the enhanced
subspace. The cross term is as cheap as the first
tangents: differentiating the vanishing residual identity $R(U(\theta),\theta)\equiv0$ twice
gives the second-order forward-sensitivity equation
\begin{widetext}
\begin{equation}
  J\,\frac{\partial^2 U}{\partial\theta_k\,\partial\theta_{k'}} \ = \ -\bigg[\,
    \frac{\partial^2 R}{\partial\theta_k\,\partial\theta_{k'}}
    \ + \ \frac{\partial^2 R}{\partial\theta_k\,\partial U}\,\frac{\partial U}{\partial\theta_{k'}}
    \ + \ \frac{\partial^2 R}{\partial\theta_{k'}\,\partial U}\,\frac{\partial U}{\partial\theta_{k}}
    \ + \ \frac{\partial^2 R}{\partial U^2}\Big[\frac{\partial U}{\partial\theta_k},\frac{\partial U}{\partial\theta_{k'}}\Big]
  \bigg],
  \label{eq:cross_tangent}
\end{equation}
\end{widetext}
with the \emph{same} Jacobian $J=\partial R/\partial U$ as the first tangent
(Eq.~\ref{eq:jacobian}), i.e.\ one further linear solve and no further nonlinear solve.
Because the spins enter the residual only through $\Ahat^2$
(Sec.~\ref{sec:param:diff}), every term on the right is a closed-form, node-diagonal
function of the source and its first two spin-derivatives, so no finite differencing
is needed. Supplying this cross term closes the off-node cross-curvature between the
two enhanced axes that a gradient-only construction leaves approximated.
Completing it is not a refinement but the
reason the enhanced pair works at all: tangents alone on two axes degrade the joint
interpolant, and it is the cross that recovers them and carries the model past the
value-only baseline. The model assembled
this way is the full-bilinear gradient-enhanced family measured in
Sec.~\ref{sec:model:joint}. Enlarging $\mathcal{E}$ beyond two axes would call for
the higher mixed partials of the correspondingly larger tensor product, and we do not
pursue it: Sec.~\ref{sec:model:enhanced} finds that a larger enhanced set degrades the
model rather than improving it.

The gradient-enhanced model is thus more accurate without any additional nonlinear
elliptic solves: it reuses derivative information already produced by the offline
build, and concentrates its accuracy gain along the slowly converging
spin directions where interpolation is hardest.
What it does cost is storage, since the tangents and their cross are kept alongside the
value at every node; Sec.~\ref{sec:model:pod} weighs the two families against stored
memory and gives the reduced-basis re-encoding that removes most of it.
Certification is unaffected: the enhanced model remains only a warm start, refined to tolerance by the certified
Newton step of Sec.~\ref{sec:polish}.

\subsection{Building the grid: warm-started continuation}
\label{sec:param:build}
The construction of the parametric model is independent of the underlying
elliptic solver. It requires only a routine
$\texttt{solve}(\theta,\mathrm{guess})$ that computes the solution at a
parameter point $\theta$ from an initial guess and, optionally, the parameter
tangent $dU/d\theta$. The interpolation nodes are populated by traversing the
tensor grid in a boustrophedon (reflected-serpentine) order, so that successive
nodes only differ by a single parameter increment. Each Newton solve can therefore
warm-start from the converged solution at a neighboring parameter point,
while only the first solve requires a cold start.

\subsection{Sparse grids for higher dimensions}
\label{sec:param:smolyak}

The full tensor-product interpolant of Eq.~\eqref{eq:surrogate} requires one
expensive elliptic solve per parameter node, so its off\-line cost $\prod_k(Q_k{+}1)$
grows exponentially with the number of parameters~$d$. This scaling is unnecessary
because the parameter-to-solution map is analytic. Analytic dependence makes the
interpolation error decay geometrically with the per-axis resolution
(Sec.~\ref{sec:param:interp}), so refining every direction to high order is
wasteful: the accuracy is dominated by a few finely resolved parameters rather than
by resolving all of them simultaneously. A \emph{Smolyak sparse
grid}~\cite{Smolyak1963,NobileTemponeWebster2008} retains only these dominant
tensor-product contributions.

Index each axis by a resolution \emph{level} $l_k$, and let $I_{l_k}$ be the
one-dimensional interpolation operator on the $m(l_k)$ nodes of that level, with
$m(0)=1$ and $m(l)=2^{l}+1$
Clenshaw--Curtis doubling nodes\footnote{The Clenshaw--Curtis nodes $\cos(\pi j/2^{l})$, $j=0,\dots,2^{l}$, are the extrema of the Chebyshev polynomials; taking $2^{l}{+}1$ of them (and thus doubling the number of subintervals each level) makes level $l$ a subset of level $l{+}1$, i.e.\ nested, since $\cos(\pi j/2^{l})=\cos(\pi(2j)/2^{l+1})$, such that the even nodes of level  $l{+}1$ are equal to the nodes of level $l$.}.
These node sets are \emph{nested}, and the
levels telescope through the hierarchical surpluses
$\Delta_{l}\equiv I_{l}-I_{l-1}$ (with $\Delta_0\equiv I_0$), so that
$I_{L}=\sum_{l\le L}\Delta_{l}$. For an analytic map the surpluses decay
geometrically, through the same Bernstein-ellipse mechanism that governs the
interpolation error (Sec.~\ref{sec:model:walls}).

In $d$ dimensions the interpolant telescopes into a sum of mixed surpluses over all
levels, $\bigotimes_k I_{L}=\sum_{\vect{l}}\bigotimes_k\Delta_{l_k}$, recovering
the $O(Q^d)$ cost. For an analytic map these mixed contributions decay geometrically
in the total level $|\vect{l}|_1=\sum_k l_k$, so resolving fine structure in
several para\-meters at once is exponentially redundant. Retaining the \emph{simplex}
$|\vect{l}|_1\le\ell$ captures nearly all of the interpolation accuracy and
discards only exponentially small contributions. The retained surpluses recombine
into the Smolyak combination technique,
\begin{equation}
  \mathcal{I}_{\rm sparse}[u] = \sum_{|\vect{l}|_1 \le \ell} c_{\vect{l}}\,
  \bigotimes_{k=1}^{d} I_{l_k}[u],
  \label{eq:smolyak}
\end{equation}
with the integer coefficients
\begin{equation}
  c_{\vect{l}} = (-1)^{\ell-|\vect{l}|_1}\binom{d-1}{\ell-|\vect{l}|_1},
  \label{eq:smolyak-coef}
\end{equation}
nonzero only for $\ell-d+1\le|\vect{l}|_1\le\ell$. Because the one-dimensional
nodes are nested, the union of these subgrids is the sparse grid, and every node is
solved only once.

The sparse grid replaces the exponential offline solve count with one that grows as $O(m\,(\log m)^{d-1})$ nodes versus $O(m^d)$,
i.e.\ near-linearly in the per-axis resolution $m$ at fixed $d$, with a
polylogarithmic factor whose order still grows with $d$: the dimensional dependence
is strongly mitigated rather than removed. It retains nearly the full
tensor-product accuracy. This scalability makes the higher-dimensional general-spin
family tractable, where the full tensor grid is not, and the saving in offline solve
count is quantified in Sec.~\ref{sec:model:joint}.

Sparse grids alter only the offline sampling strategy: each subgrid interpolant in
Eq.~\eqref{eq:smolyak} is itself an instance of the parametric model of
Eq.~\eqref{eq:surrogate}, so the online interpolation, the certified Newton
refinement of Sec.~\ref{sec:polish}, and the differentiability of
Sec.~\ref{sec:param:diff} carry over unchanged. When the parameters differ in
difficulty, as the slowly converging spin axes of Sec.~\ref{sec:model:peraxis} do,
the isotropic simplex can be replaced by a \emph{dimension-adaptive} downward-closed
index set~\cite{NobileTemponeWebster2008} grown greedily along the axes whose
surpluses remain large, a refinement we leave to future work. In every case the
sparse grid reduces the number of expensive offline solves without weakening the
certification guarantee or the analytic differentiability that define the method.

\subsection{Reduced-basis re-encoding}
\label{sec:param:pod}
Because the model \emph{is} its stored nodal solves (Eq.~\ref{eq:surrogate}), its
memory grows linearly in the node count, which becomes a liability in
high dimension. However, since the parameter-to-field map is smoothly low
rank, the corpus admits a low-rank reduced-basis re-encoding.

We compress by \emph{proper orthogonal decomposition} (POD, equivalent to
principal-component analysis)~\cite{Sirovich1987,BerkoozHolmesLumley1993}. Stacking
the $N$ stored fields as rows of a matrix and subtracting their mean $\bar u$, a
singular-value decomposition yields orthonormal \emph{spatial} modes
$\Phi=[\phi_1,\phi_2,\dots]$ ordered by singular value $\sigma_k$; each stored field
can then be written as $u=\bar u + \Phi\,c$ with coefficients $c=\Phi^{\!\top}(u-\bar u)$. The
singular values decay exponentially, because the map is analytic in $\theta$, the
same analyticity that governs the interpolation error (Sec.~\ref{sec:model:walls}).
Keeping the leading $r$ modes, we store $\Phi$ together with the length-$r$
coefficient vector at each node, i.e.\ $r$ numbers per node instead of the full field, and, at query time, interpolate the \emph{coefficients} with the identical
barycentric/Smolyak machinery of Eq.~\eqref{eq:surrogate} and decode
$u(\theta)=\bar u+\Phi\,c(\theta)$.

Two properties keep the truncation harmless. First,
interpolation is linear, so interpolating the projected coefficients is
\emph{identical} to projecting the full-field interpolant onto the $r$ leading
modes: the compressed model is exactly the rank-$r$ truncation of
Eq.~\eqref{eq:surrogate}, differing from it only by a controllable truncation tail.
Second, and this is why even that tail is harmless, the certified polish of
Sec.~\ref{sec:polish}
reconstructs precisely the high-derivative content the truncation discards. The differentiable evaluation of
Sec.~\ref{sec:param:diff} also survives: the parameter Jacobian of the compressed
model is the full Jacobian projected onto $\Phi$, which loses a negligible fraction
of the gradient. The decomposition is computed once from the existing corpus, with no
new solves; Sec.~\ref{sec:model:pod} reports the compression achieved on the quasi-circular demonstration
models.

\section{Quasi-circular parametric models}
\label{sec:model}
We now assemble the two headline models, both \emph{quasi-circular} orbiting
initial data built by the identical machinery over the production box
$b\in[3,10]\,M$ (i.e.\ puncture separations $D=2b\in[6,20]\,M$), $q\in[1,3]$, and
dimensionless spins with every component in $[-0.9,0.9]$, at fixed total mass
$M=1$: a four-dimensional \emph{aligned-spin} model
$\theta=(b,q,\chi^{A}_{y},\chi^{B}_{y})$ and the full eight-dimensional
\emph{general-spin} model $\theta=(b,q,\boldsymbol{\chi}^{A},\boldsymbol{\chi}^{B})$,
whose in-plane spin axes are centered on zero so the
aligned model nests, node-for-node, inside it.
Table~\ref{tab:box} collects the box and the build configuration the two models
share, for reference. The spin box is a hypercube in the components, so its corners
combine them into $|\boldsymbol{\chi}^{X}|=0.9\sqrt3\approx1.56$; note these are
Bowen--York free-data parameters rather than horizon spins.

\begin{table}[t]
  \caption{\label{tab:box}%
    \textbf{Parameter box and build configuration} shared by the two
    quasi-circular models (Sec.~\ref{sec:model}). Both models sample the same edges on
    the same spatial grid at the same level: the aligned-spin model varies only
    $\chi^{A}_{y}$ and $\chi^{B}_{y}$, the general-spin model varies all six spin components. The
    momenta follow from the other parameters through the
    quasi-circularity condition (Sec.~\ref{sec:model:qc}).}
\begin{ruledtabular}
\begin{tabular}{lc}
  puncture separation $D=2b$ $[M]$ & $[6,\,20]$ \\
  mass ratio $q=m_A/m_B$ & $[1,\,3]$ \\
  spin components $\chi^{A}_{i},\chi^{B}_{i}$ & $[-0.9,\,0.9]$ \\
  total bare mass $M=m_A+m_B$ & $1$ \\
  Smolyak level $\ell$ & $5$ \\
  spatial grid $(N_A,N_B,N_\phi)$ & $(44,32,8)$ \\
  gradient-enhanced axes $\mathcal{E}$ & $\{\chi^{A}_{y},\chi^{B}_{y}\}$ \\
\end{tabular}
\end{ruledtabular}

\end{table}

The accuracy metric throughout is the \emph{held-out} interpolation error: at a
parameter point that is \emph{not} a collocation node we compare the interpolant
to a direct elliptic solve on the \emph{same} frozen spatial grid ($N_A=44$,
$N_B=32$, $N_\phi=8$) and take the largest nodal difference between the two
fields. Because both are represented on the same grid, the spatial
discretization error cancels and what remains is the pure
parameter-interpolation error. We report it in two ways. The \emph{per-axis}
error sweeps a single parameter through $21$ off-node points with the other axes
held at a random base point, and keeps the worst of the $21$; repeating that sweep
at $100$ random base points turns it into a distribution, and isolates the
convergence rate of the swept axis (Fig.~\ref{fig:peraxis}, and
Fig.~\ref{fig:walls} at fixed rather than random base points). The \emph{joint}
held-out error instead draws $1000$ parameter
points uniformly at random from the whole box, so that every axis is off-node at
once; it is the error encountered at an arbitrary query, and it is governed by
the slowest-converging axis (Fig.~\ref{fig:joint}). Both are distributions over the sampled points,
which we report as a median with its extremes.

\subsection{The quasi-circular momentum condition}
\label{sec:model:qc}
What makes the family \emph{orbiting} rather than a free momentum scan is that the
per-puncture momenta are not free parameters but a deterministic function of
$(b,q,\text{spins})$, fixed by a post-Newtonian quasi-circularity condition: a
closed-form tangential momentum together with a small radial component, so every node is a physical orbiting configuration.
The tangential momentum lies along $x$, off
the separation ($z$) axis, so the orbital angular momentum points along $+y$ and
the data are non-axisymmetric. We fix the momenta by the standard
puncture quasi-circularity procedure~\cite{Husa2008,Walther2009}. Defining
$M \equiv m_A+m_B$, $\mu \equiv m_Am_B/M$, $\nu \equiv \mu/M$, coordinate separation $D \equiv 2b$, and
$x\equiv M/D$, the per-puncture \emph{tangential} momentum is the non-spinning
post-Newtonian series through 3PN~\cite{Walther2009},
\begin{equation}
\begin{aligned}
    \frac{P_{\rm t}}{\mu} \, &= \, x^{1/2} + 2\,x^{3/2} + \tfrac{1}{16}(42-43\nu)\,x^{5/2} \\
        &\quad + \tfrac{1}{128}\bigl(480 + (163\pi^2-4556)\nu + 104\nu^2\bigr)\,x^{7/2} .
\end{aligned}
\label{eq:pt}
\end{equation}
Note that the leading term is the Newtonian value $P_{\rm t}\to\mu\sqrt{M/D}=\mu\sqrt{M/2b}$. We
add the leading (1.5PN) spin--orbit
correction~\cite{Healy2024}, which depends only on the spin projections $\chi^{X}_{\parallel}$
onto $\hat{\vect{L}}$; in-plane spin components first enter at higher PN order and
are neglected. Converted to the convention used throughout this
paper\footnote{Ciarfella \emph{et al.}~\cite{Healy2024} write this correction with
$q=m_2/m_1\le1$ (body $1$ the larger hole); Eq.~\eqref{eq:ptso} is the
algebraically identical form obtained by substituting $q\to1/q$ and relabeling
$(1,2)\to(A,B)$. It is invariant under $q\to1/q$ with
$\chi^{A}_{\parallel}\leftrightarrow\chi^{B}_{\parallel}$.}
($q=m_A/m_B\ge1$, $A$ the larger hole; $\chi^{X}_{\parallel} \equiv (\vect{S}^{X}\!\cdot\!\hat{\vect{L}})/m_X^2$
the dimensionless spin projected on the orbital angular momentum, i.e.\ the aligned component), it reads
\begin{equation}
  \frac{P_{\rm t}^{\rm SO}}{\mu} \, = \, -\frac{2}{3(1+q)^2}\Bigl(q(4q+3)\,\chi^{A}_{\parallel} \hspace{0.07em} + \hspace{0.07em} (3q+4)\,\chi^{B}_{\parallel}\Bigr)\,x^2 .
  \label{eq:ptso}
\end{equation}
The tangential momentum that the data actually carry is the sum of the two,
\begin{equation}
  P_{\rm t}^{\rm tot} \ = \ P_{\rm t}^{\vphantom{\rm tot}} \, + \, P_{\rm t}^{\rm SO},
  \label{eq:pttot}
\end{equation}
which reduces to Eq.~\eqref{eq:pt} whenever both spins are non-spinning or purely
in-plane.
The \emph{radial} component is the leading radiation-reaction (Peters) momentum
$P_{\rm r}=\mu\,|\dot D|$ with the quadrupole inspiral rate
$\dot D=-\tfrac{64}{5}\mu M^2/D^3$~\cite{Peters1964,Walther2009}, i.e.
\begin{equation}
  \frac{P_{\rm r}}{\mu} \ = \ \frac{64}{5}\,\frac{\mu M^2}{D^3},
  \label{eq:pr}
\end{equation}
so $P_{\rm r}/P_{\rm t}\propto\nu\,(M/D)^{5/2}\!\to0$ as $b\to\infty$, i.e\ the circular limit.
The per-puncture momenta are then $\vect{P}_A=(+P_{\rm t}^{\rm tot},0,-P_{\rm r}^{\vphantom{\rm tot}})$,
$\vect{P}_B=(-P_{\rm t}^{\rm tot},0,+P_{\rm r}^{\vphantom{\rm tot}})$, with zero net linear momentum and orbital angular
momentum $2bP_{\rm t}^{\rm tot}$ along $+y$; only the aligned ($S_y$) spin component enters
Eq.~\eqref{eq:ptso}, so in-plane spins leave the momenta unchanged at this order.
The angular momentum this produces is validated against its closed form and against
Two\-Punctures, and the large-separation Newtonian
limit of Eq.~\eqref{eq:pt} is the cross-check that the two codes share a momentum and
separation convention (Appendix~\ref{sec:validation:qc}).

\subsection{Convergence across the parameter space}
\label{sec:model:peraxis}
\begin{figure*}[t]
  \centering
  \includegraphics[width=\textwidth]{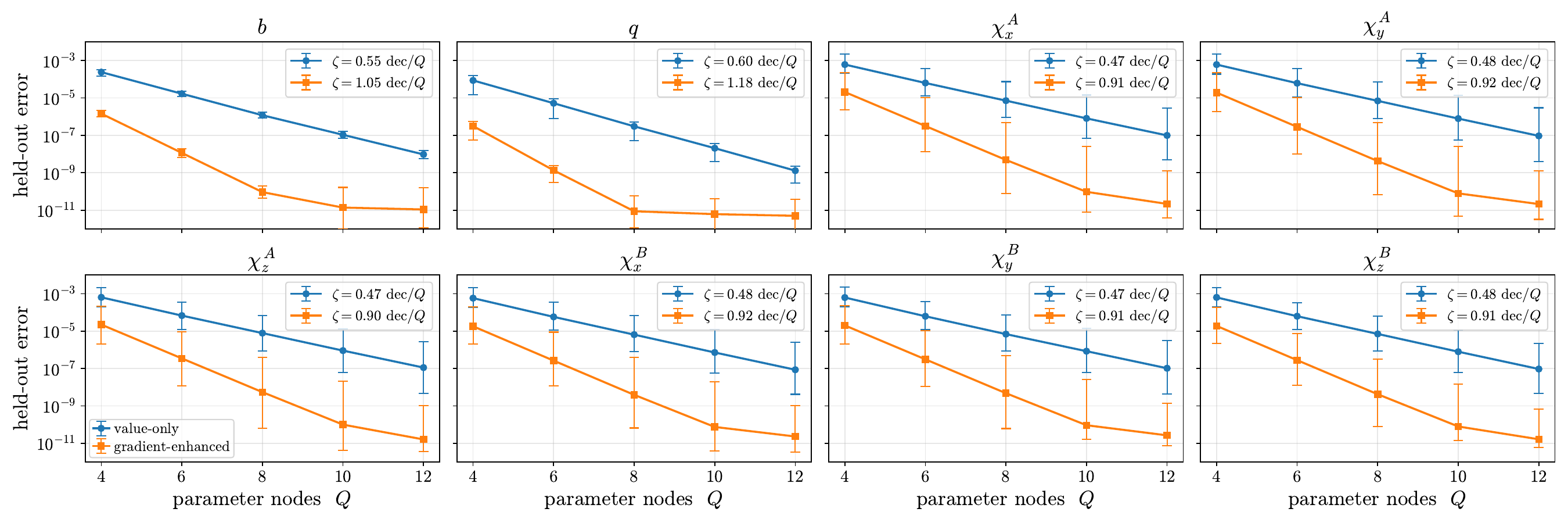}
  \caption{\label{fig:peraxis}%
    \textbf{Per-axis convergence of the quasi-circular models.}
    Held-out interpolation error against the number of parameter nodes for each of
    the eight axes $(b,q,\boldsymbol{\chi}^{A},\boldsymbol{\chi}^{B})$; the first four
    are the four-dimensional aligned-spin model. The value-only model is
    compared with the gradient-enhanced one.
    A one-parameter sweep carries no mixed tangent, so the cross term of
    Eq.~\eqref{eq:hermite} does not enter
    here. Each plotted error is the worst over $21$ off-node points along the swept
    axis; markers are medians of that worst case over $100$
    random base points for the non-swept axes, with whiskers spanning its
    best--worst range across those base points. Every axis decays geometrically and matching the tangent roughly doubles each rate.}
\end{figure*}

Sweeping one parameter at a time, every axis converges exponentially
(Fig.~\ref{fig:peraxis}). The separation $b$ and mass ratio $q$ are the fastest;
the spins are somewhat slower, but all eight rates lie within a narrow band, and
the two holes' spin axes converge at comparable rates. Gradient enhancement (Sec.~\ref{sec:param:hermite}) roughly doubles every
rate for the price of one extra linear solve per node against the node's
Jacobian. Because the joint held-out error is governed by the
slowest of these axes, it is the spin directions that set the offline cost of the
model (Fig.~\ref{fig:joint} and Sec.~\ref{sec:model:joint}).

\subsection{Analyticity limits of the parameter map}
\label{sec:model:walls}

The convergence rate of a spectral parameter interpolant is set by the distance
from the sampling interval to the nearest singularity of the parameter-to-solution
map in the complex plane, through the \emph{Bernstein-ellipse} convergence bound
(Theorem~8.2 of Ref.~\cite{Trefethen2019}, going back to \cite{Bernstein1912}).
We map the sampling interval $[\theta_{\min},\theta_{\max}]$
of a parameter $\theta$ affinely onto $[-1,1]$ by
$\theta\mapsto\xi=(2\theta-\theta_{\max}-\theta_{\min})/(\theta_{\max}-\theta_{\min})$,
and consider the \emph{Bernstein ellipses} $E_\varrho$, which are defined as the ellipses in the complex
$\xi$-plane with foci at $\pm1$ whose semi-major and semi-minor axes sum to
$\varrho>1$. If the parameter-to-solution map is analytic inside $E_\varrho$, its $Q$-node Chebyshev interpolant
converges exponentially, with error $\sim\varrho^{-Q}$, i.e.\ a rate of $\log_{10}\varrho$
decades per node. The operative $\varrho$ is fixed by the \emph{largest} ellipse free
of singularities, so the nearest singularity of the map lies \emph{on} $E_\varrho$. A
singularity at the mapped point $\xi_\ast$ sits on the ellipse of parameter
$\varrho=|\xi_\ast+\sqrt{(\xi_\ast)^2-1}|$, which for a real singularity outside the
interval reduces to $\varrho=|\xi_\ast|+\sqrt{(\xi_\ast)^2-1}$. Hence, a singularity in parameter space close to the
sampled box $[\theta_{\min},\theta_{\max}]$ will yield a $\varrho$ just above unity and thus slow convergence, while a distant singularity will yield a
large $\varrho$ and fast convergence.

\begin{figure}[!t]
  \centering
  \includegraphics[width=\columnwidth]{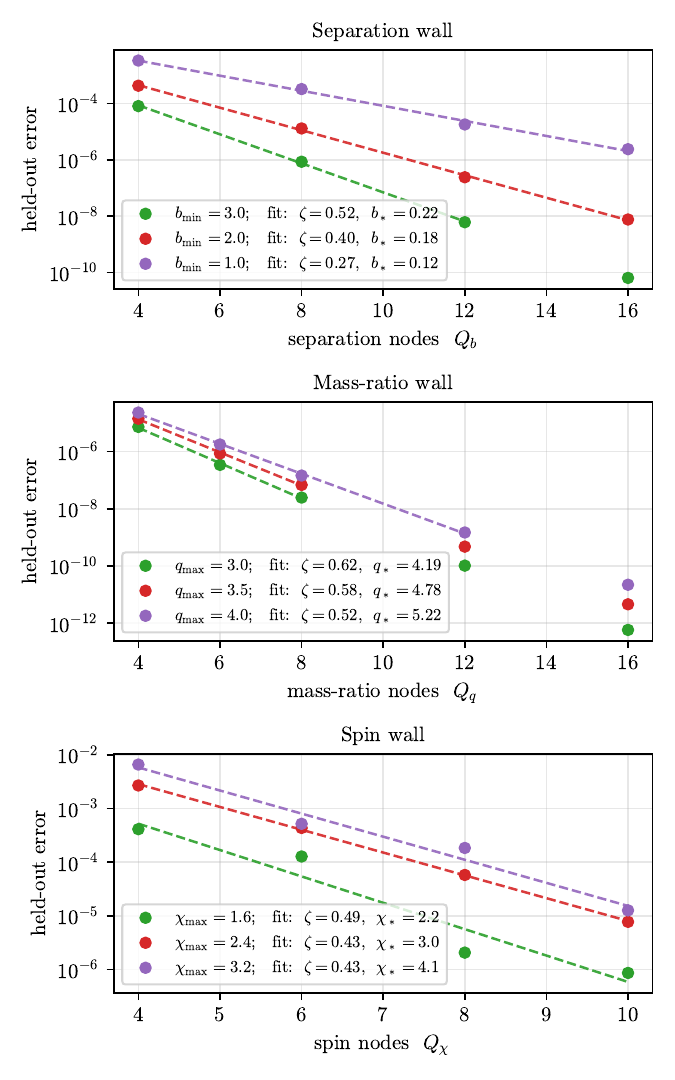}
  \caption{\label{fig:walls}%
    \textbf{Analyticity walls of the quasi-circular family.} Held-out
    interpolation error against the number of nodes on the swept axis, for several
    sampling ranges, with the exponential fit $\varepsilon\approx A\,10^{-\zeta Q}$
    dashed over the fitted window, which ends where the error saturates on the
    float64 floor. Legends give, per range, the fitted rate $\zeta$ (decades/node)
    and the inferred nearest real singularity $\theta_\ast$.
    \emph{Top (separation $b$):} the rate degrades as the interval is pushed toward
    coincidence while $b_\ast$ stays around $b_\ast \approx 0$, i.e.\ a hard, real wall.
    \emph{Middle (mass ratio $q$):} the rate degrades only mildly and $q_\ast$
    recedes roughly in proportion to the interval, so the box never approaches it, i.e.\ a complex pair, hence soft.
    \emph{Bottom (spin $\chi$):} the rate degrades only mildly and $\chi_\ast$
    marches outward, beyond the sampled range, i.e. again a complex-conjugate pair off the
    real axis, hence soft.}
\end{figure}

We read this estimate \emph{backwards} to locate the wall. From the slope of the
held-out error against the node count $Q$ we measure the geometric convergence rate
$\zeta \equiv \log_{10}\varrho$, invert it to
$|\xi_\ast|=\tfrac12(\varrho+\varrho^{-1})$, and map $\xi_\ast$ back through the affine map, taking the real root outside the sampled box on the physically singular side
(small $b$ toward merger for the separation; large $q$ toward the test-mass limit
for the mass ratio; large $|\chi|$ toward the extremal
limit for the spins), to read
off the inferred singularity location $\theta_\ast$. Sweeping the sampling range
then diagnoses the \emph{character} of the wall. A genuine real branch point stays
pinned at a fixed $\theta_\ast$ as the range changes. A complex-conjugate pair off
the real axis has no real singularity at all, yet the one-parameter fit still
returns a real $\theta_\ast$, the real point lying on the same Bernstein ellipse
as the true complex pair, which \emph{marches} outward as the range grows. The
legends of Fig.~\ref{fig:walls} report this inferred $\theta_\ast$ for each sampling
range alongside the fitted rate; whether $\theta_\ast$ holds still or recedes as the
range grows separates the two wall types below.

The quasi-circular family has three walls, one hard and two soft
(Fig.~\ref{fig:walls}). The separation and spin sweeps are run at equal mass with
the remaining parameters at their symmetric values; the mass-ratio sweep, whose
swept axis is itself the mass ratio, instead holds $b=3.5\,M$ and both spins at
zero.

The \emph{merger} wall, in the separation $b$, is hard. As the sampling interval
is pushed toward coincidence the rate degrades smoothly and tracks the Bernstein
prediction for a singularity fixed at $b=0$, the inferred $b_\ast$ staying at the
merger throughout. This resolves the concern (specific to the orbiting family) that the
$b\to0$ limit is \emph{doubly} singular: the punctures coincide \emph{and} the
post-Newtonian tangential momentum $P_{\rm t}\propto b^{-1/2}$ diverges, yet the
extra branch does not \emph{relocate} the wall off $b=0$. Its only signature is a
mild undershoot of the measured rate below the pure-coincidence prediction at the
smallest $b_{\min}$, i.e.\ the momentum divergence makes the wall marginally
\emph{stronger}, not closer.

The \emph{mass-ratio} wall is soft. Widening the sampled interval degrades the rate
only mildly, from $0.62$ to $0.52$ decades per node between $q_{\max}=3$ and
$q_{\max}=4$, while the inferred nearest real singularity recedes from
$q_\ast=4.19$ to $5.22$, holding $q_\ast/q_{\max}$ near $1.3$--$1.4$ throughout. A
feature that retreats in proportion to the interval is a complex-conjugate pair
rather than a real branch point.

The \emph{spin} wall, in $\chi$, is also soft. The measured rate degrades
only mildly as the range grows. Here, the range was deliberately extended past the production
box to locate the wall and the inferred nearest \emph{real} singularity
marches outward, staying beyond the sampled range throughout, which is the
signature of a complex-conjugate pair off the real axis rather than of a real
wall. That nearest feature is already super-extremal, well past the Cook--York
horizon-spin ceiling~\cite{CookYork1990}, so spin is a benign axis with no sharp
singularity near the physical box. What sets the convergence rate is therefore
the \emph{distance} of the nearest singularity and not its hardness: the family's
only hard wall lies on one of its two fastest axes, while its slowest axes carry
only soft, receding ones.

\subsection{Cost, memory, and certified residuals}
\label{sec:model:joint}

\begin{figure*}[t]
  \centering
  \includegraphics[width=\textwidth]{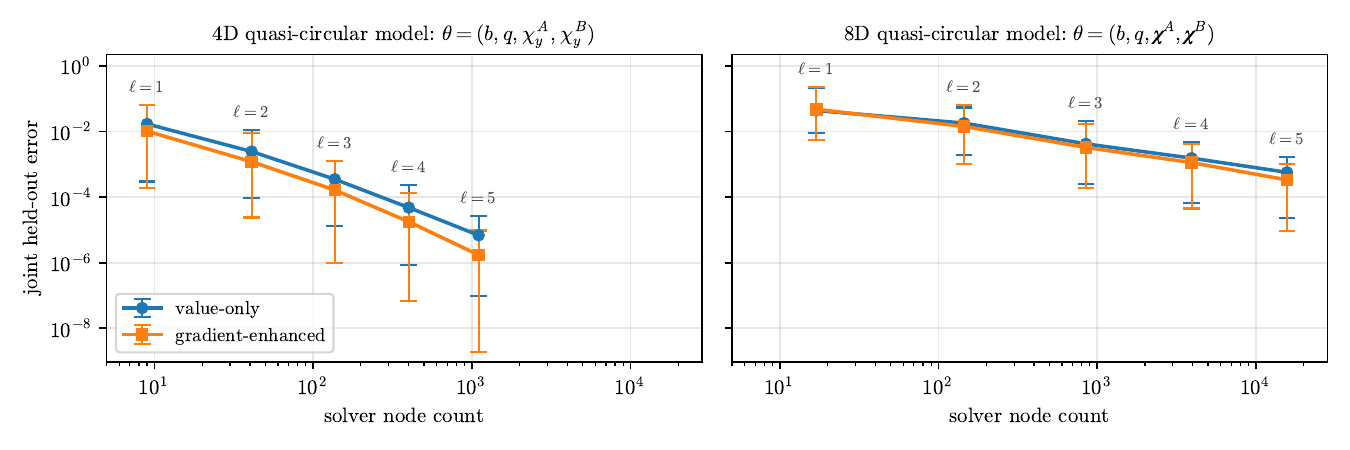}
  \caption{\label{fig:joint}%
    \textbf{Joint held-out convergence.} Joint held-out error over $1000$ random
    off-node points against solver node count (logarithmic, shared between
    panels), with the Smolyak level $\ell=1$--$5$ labeled at each point, shown as
    the best, median, and worst of the distribution.
    \emph{Left (four-dimensional model):} the
    value-only sparse interpolant against the gradient-enhanced
    model enhanced on the two spin axes, which is more accurate
    at every level.
    \emph{Right (eight-dimensional model):} the same two curves.}
\end{figure*}

The offline cost of a full tensor grid scales as $O(Q^d)$; a
sparse (Smolyak) grid~\cite{Smolyak1963,NobileTemponeWebster2008} breaks this and is
what makes both the four-dimensional and the eight-dimensional quasi-circular
models tractable. Figure~\ref{fig:joint} shows that cost
curve: joint held-out accuracy against the number of elliptic solves that bought it,
level by level. At isotropic Smolyak level $\ell=5$ the four-dimensional model is
$1105$ solver nodes and the eight-dimensional one $15{,}713$, and the nested lower
levels $\ell=1\subset\cdots\subset5$ are fully contained in each corpus, so no lower
level costs additional solves. A full tensor grid at the four-dimensional sparse
grid's finest level ($33$ nodes per axis) would require $\sim\!10^3 \times$ more nodes.

\begin{figure*}[t]
  \centering
  \includegraphics[width=\textwidth]{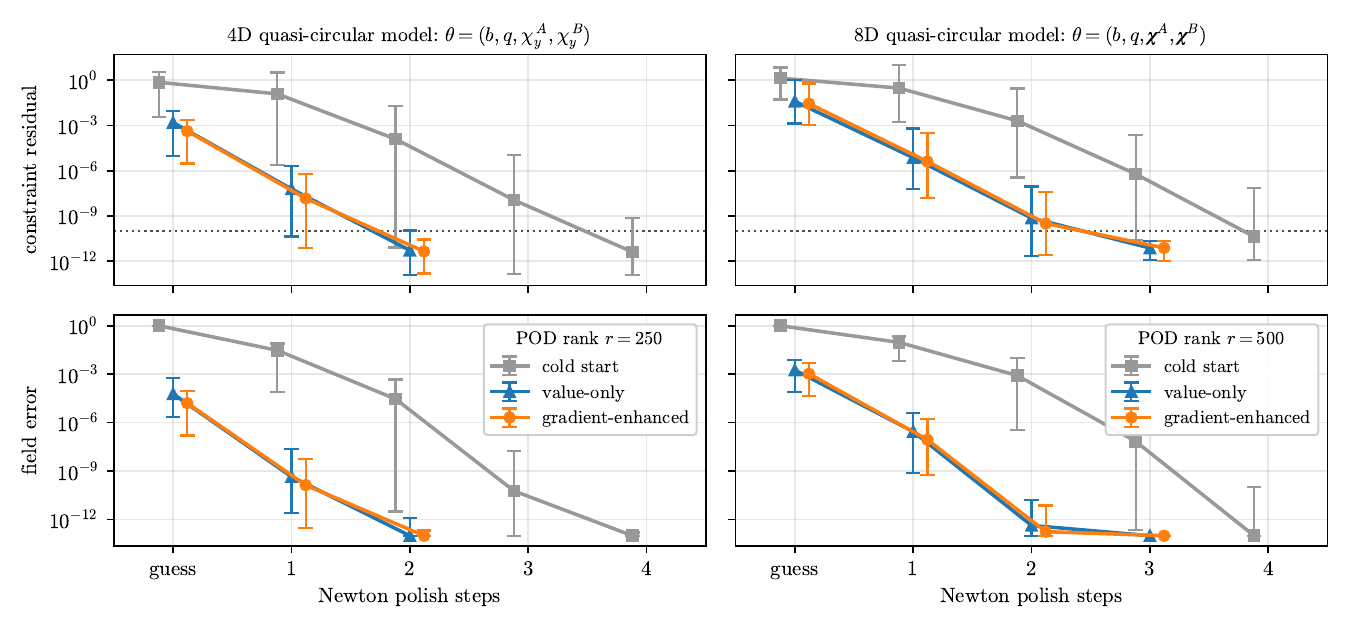}
  \caption{\label{fig:polish}%
    \textbf{Certified refinement.} Row-equilibrated constraint residual $\|R\|^{(v)}_\infty$ (top row)
    and relative field error $\|U-U_{\rm true}\|_2/\|U_{\rm true}\|_2$ (bottom
    row) over $1000$ uniformly-random off-node points of the quasi-circular models
    (left column four-dimensional
    aligned-spin, right column eight-dimensional general-spin), before the polish
    (``guess'') and after each Newton step. Each curve is the median with
    min--max whiskers over the $1000$ points, the field error measured against the
    converged certified iterate. \emph{Dotted:} the certification threshold
    $10^{-10}$ in the row-equilibrated norm $\|R\|^{(v)}_\infty$ of
    Eq.~\eqref{eq:equil_residual}. Both value-only and gradient-enhanced models are POD-compressed onto the
    same reduced basis at the same rank. In both dimensions, both
    warm starts certify \emph{every} point to $\|R\|^{(v)}_\infty\le10^{-10}$ in at most
    three Newton steps.}
\end{figure*}

Evaluating the interpolant is cheap. A bare query is a single contraction over
the stored nodal solves, with no Newton iteration and no factorization, and its
cost is dominated by streaming those stored fields; the reduced-basis
re-encoding of Sec.~\ref{sec:model:pod} is therefore what makes it small. A query costs a fraction of a Newton--Krylov step of the
elliptic solve, so the warm start is essentially free relative to the solve it
seeds. A certified query is that same warm-started solve, and its saving over a cold solve
is the reduced step count: two to three polish steps rather than the four to five a
cold start needs (Fig.\ \ref{fig:polish}).
Because the model is nothing but its stored nodal solves, its memory is
\emph{linear} in the node count.

Two implementation choices are what make a query that cheap.
Rather than evaluate the combination one subgrid
at a time, we distribute the sum through each subgrid's linearity, which turns a
query into a few matrix--vector products against the deduplicated node pool;
the subgrids overlap heavily, the eight-dimensional model's $15{,}713$ distinct
nodes filling $101{,}575$ subgrid slots. The evaluator is then compiled once per
model. Neither choice changes the interpolant, but both reorder its
floating-point summation, so a query differs from the direct subgrid-by-subgrid
sum by less than $10^{-14}$ relative (far below the interpolation error).
We retain the direct sum in the released code as
the reference implementation against which the default path is tested.

\subsection{How many axes to gradient-enhance?}
\label{sec:model:enhanced}

We tested whether the gradient-enhanced subspace $\mathcal{E}$ should be enlarged beyond the two aligned-spin axes. It should
not. Enhancing a \emph{single} axis always improves on the value interpolant, but
enhancing several axes with tangents alone degrades it, and it does so whichever axes
are chosen. Extending $\mathcal{E}$ to all four axes of the aligned-spin box at the same
Smolyak level costs two orders of magnitude in held-out median relative to enhancing
nothing at all, and completing the pairwise-bilinear crosses does not repair that. The
two-axis case is the one that can be rescued: with tangents alone the aligned-spin pair
is also worse than value-only, and it is the bilinear cross term that improves it beyond the value-only model.

The practical consequence is that $\mathcal{E}$ should be kept small and spent on the
slowest axes, and that the shipped $\mathcal{E}=\{\chi^{A}_{y},\chi^{B}_{y}\}$,
completed to full bilinear, is at or near the optimum of the configurations we
measured.

\subsection{Reduced-basis compression of the model}
\label{sec:model:pod}

The corpora are correspondingly large: the gradient-enhanced four-dimensional
$\ell=5$ corpus (values, the two enhanced-axis node tangents, \emph{and} their
bilinear cross tangent) is $388$~MiB and the eight-dimensional one $5.4$~GiB.
This is inconveniently large, and it is entirely removable by the reduced-basis
(POD) re-encoding of Sec.~\ref{sec:param:pod}. For the four-dimensional model the rank-$r$
reconstruction error of the corpus falls below $10^{-6}$ by $r=87$ modes, and
because the value and its parameter tangents \emph{share} one spatial basis, the
enhanced corpus needs only some $7\%$ more modes than the value-only one at the
same tail. The rank at a fixed accuracy is set by the intrinsic dimension of the
solution manifold, not by the sampling density: it is the counterpart to
Sec.~\ref{sec:param:smolyak} along the orthogonal axis, i.e.\ the Smolyak level fixes
the parameter resolution and the offline solve count, POD the spatial rank, and
the two are independent.

\begin{figure*}[t]
  \centering
  \includegraphics[width=\textwidth]{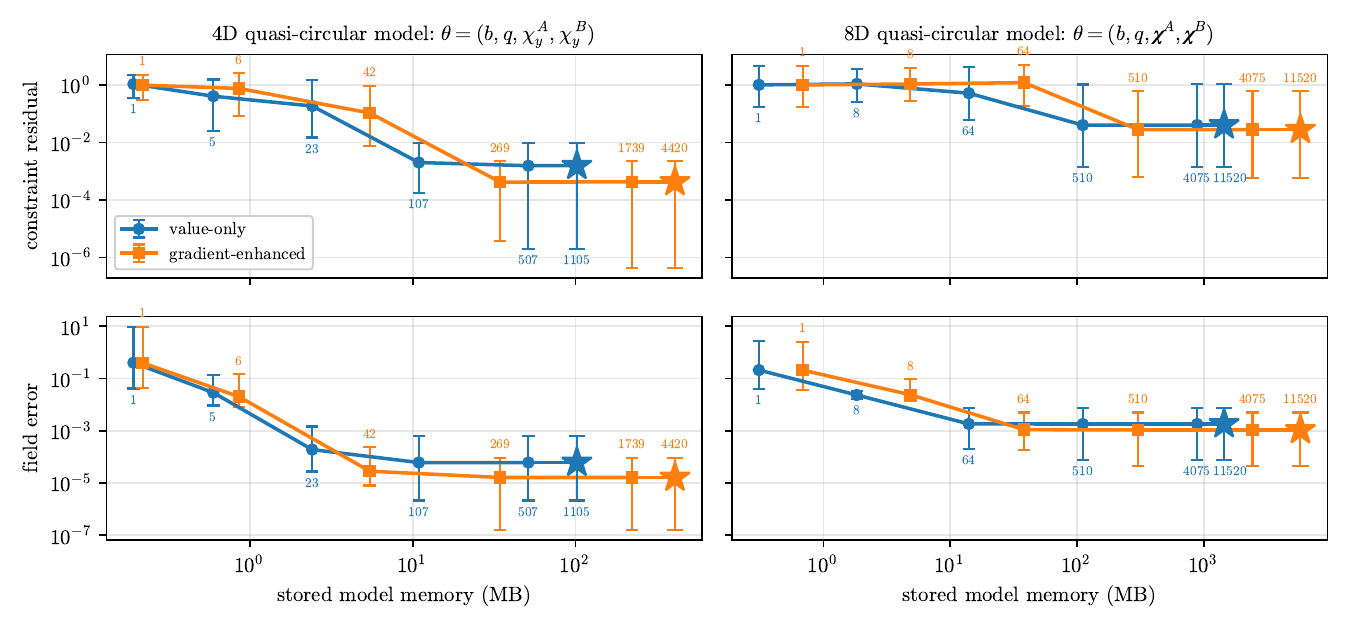}
  \caption{\label{fig:pod}%
    \textbf{Reduced-basis (POD) compression versus stored memory.} Constraint
    residual $\|R\|^{(v)}_\infty$ (top row) and
    relative field error $\|U-U_{\rm true}\|_2/\|U_{\rm true}\|_2$ (bottom row), as the POD
    truncation rank $r$ is swept. Median line with min--max whiskers over $1000$
    uniformly-random off-node query points; each point is labeled by $r$ and
    $\star$ marks the full-rank model.
    Columns share the memory axis:
    \emph{left,} the four-dimensional and \emph{right,} the
    eight-dimensional quasi-circular model.
    Compressing raises the residual, but this is fixed by the polishing Newton steps (Fig.~\ref{fig:polish}).}
\end{figure*}

Figure~\ref{fig:pod} shows the resulting memory--accuracy tradeoff. At the shipped
ranks ($r=250$ and $r=500$) the four-dimensional model shrinks by a factor of $13$
($388$~MiB to $30$~MiB) and the eight-dimensional one by $19$ ($5.4$~GiB to $284$~MiB), and
per-query evaluation, now a barycentric sum over $r$ coefficients rather than the
full field, drops to a few milliseconds. The certified constraint residual, the
number of Newton steps needed to reach it (Sec.~\ref{sec:polish}), and the
verified $\dID$ (Sec.~\ref{sec:param:diff}) are all preserved.

\subsection{External validation}
\label{sec:model:validation}
The underlying initial data reproduce the established
TwoPunctures~\cite{AnsorgBruegmannTichy2004} solver to spectral accuracy on the
most demanding, genuinely non-axisymmetric configurations, and they are
evolution-ready: on an evolution grid their finite-difference constraint violation
is indistinguishable from TwoPunctures-sourced data. Appendix~\ref{sec:validation} gives the full
comparison.

\section{Applications}
\label{sec:applications}
The two payoffs of the method are that it is differentiable and
constraint-certified, and we
demonstrate both in two applications: certified parameter targeting on a
four-dimensional quasi-circular model, and eccentricity reduction on a
purpose-built two-dimensional $(b,P_{\rm t})$ model. Each targets
a standard numerical-relativity initial-data workflow, and in each the honest cost
metric is the number of \emph{certified} elliptic solves, i.e.\ full nonlinear solves
whose returned field has passed the $\|R\|^{(v)}_\infty\le10^{-10}$ gate. This is the unit used in every cost statement and figure
below; the Newton steps and linear solves inside one such solve are not counted
separately.

Both applications read physical observables off that certified field. The
Arnowitt--Deser--Misner (ADM) mass is the monopole of the solved conformal factor,
$\Madm=(m_A+m_B)-2b\,\langle\partial_A u\rangle_{A=1}$ on the $m=0$ mode at the
compactified outer edge~\cite{BaumgarteShapiro2010}; the angular momentum $J$ is the Bowen--York surface
integral, which for the nonspinning configurations targeted here reduces to the
closed form $2bP_{\rm t}^{\rm tot}$ of Sec.~\ref{sec:model:qc}; and the individual hole
masses are the Brandt--Br\"ugmann puncture masses~\cite{BrandtBruegmann1997}
$M_X=m_X\bigl(1+u(\vect{x}_X)+m_Y/2D\bigr)$ with $Y\neq X$, evaluated at the
punctures on the same field. Each is therefore a differentiable function of the
parameters through $u$, and each is validated against TwoPunctures in
Appendix~\ref{sec:validation:qc}.

\subsection{Certified parameter targeting}
\label{sec:applications:targeting}

The first workflow is parameter control~\cite{Mendes2025}: find the free data that
realize a chosen physical configuration. The standard tool is a black-box control
loop, i.e.\ an outer Newton/Broyden iteration that adjusts the free data and performs a
full constraint solve at every step to measure the resulting physical parameters. We
target the ADM mass and angular momentum $(\Madm,J)$ by adjusting $(b,q)$ at zero
spin, over $100$ random known-answer targets, and compare three strategies, all on
the four-dimensional value-only sparse-grid model of Sec.~\ref{sec:param:smolyak},
on which the targeting loop is model-agnostic. A \emph{cold} black box
solves the constraints from scratch at each control evaluation. A \emph{warm} black
box seeds each solve from the interpolant, so every constraint solve begins close to
its answer. The \emph{differentiable} method runs the entire Gauss--Newton control
loop on the parametric model using the analytic $\partial F/\partial\theta$, so
no elliptic solve and no finite-difference Jacobian enters the loop. It then
certifies the result and, because the free model carries interpolation error, applies
a short last-mile correction whose every step is re-certified, so the target residual
is always measured on a certified field. That certification and its last-mile
correction are the certified solves counted below.

\begin{figure}[t]
  \centering
  \includegraphics[width=\columnwidth]{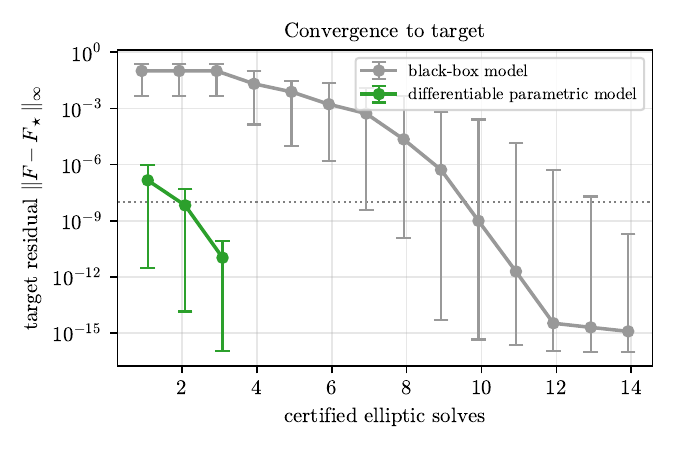}
  \caption{\label{fig:targeting}%
    \textbf{Certified parameter targeting.} Reaching a physical target
    $(\Madm,J)$ by adjusting $(b,q)$, over $100$ random known-answer targets,
    measured in \emph{certified} elliptic solves. Each method is drawn as the
    across-target median residual at each solve count, with min--max whiskers; every
    target is carried to a common solve count, so each point is the full set of
    $100$. Each curve ends where its last target reaches the control tolerance
    (dotted): three certified solves for the differentiable parametric model, with a
    median of two, against fourteen for the black-box loop, with a median of ten.}
\end{figure}

Differentiability changes the cost class here
(Fig.~\ref{fig:targeting}). The gradient method reaches every target in at most
three certified solves, with a median of two, whereas both black-box variants take a
target-dependent seven to fourteen, with a median of ten.
Warm-starting cuts the per-solve cost but \emph{not}
the number of solves: the outer control iteration is unchanged, so a warm start alone
never leaves the black box's complexity class. The distinction is structural. A
black-box loop can drive the residual to zero only through function evaluations, and
each one is a certified elliptic solve; a low-rank Jacobian update (Broyden's method,
which sidesteps the $d$ extra solves a finite-difference Jacobian would cost~\cite{Mendes2025}) lowers
the iteration count but cannot remove the one solve per iteration. The analytic
gradient removes those per-iteration solves outright, running the search entirely on
the parametric model and leaving only the last-mile certification.
This saving, moreover, widens with the number of controlled parameters.

\subsection{Eccentricity reduction}
\label{sec:applications:ecc}

The second workflow is eccentricity control, the most common tuning step in
preparing quasi-circular data. We use the Cook effective-potential
method~\cite{Cook1994}: along a sequence of fixed angular momentum $J$ (radial
momentum $P_{\rm r}=0$, i.e.\ an apsis), the binding energy $E_{\rm b}=\Madm-(M_A+M_B)$ has a minimum at the circular orbit
$\partial E_{\rm b}/\partial b|_J=0$, and a mildly off-circular apsis at $(b_0,J)$ has
eccentricity $e=|b_0-b'|/(b_0+b')$ set by the second turning point $b'$ across that
minimum. Reducing eccentricity is driving $b_0\to b_{\rm circ}$.

\begin{figure}[t]
  \centering
  \includegraphics[width=\columnwidth]{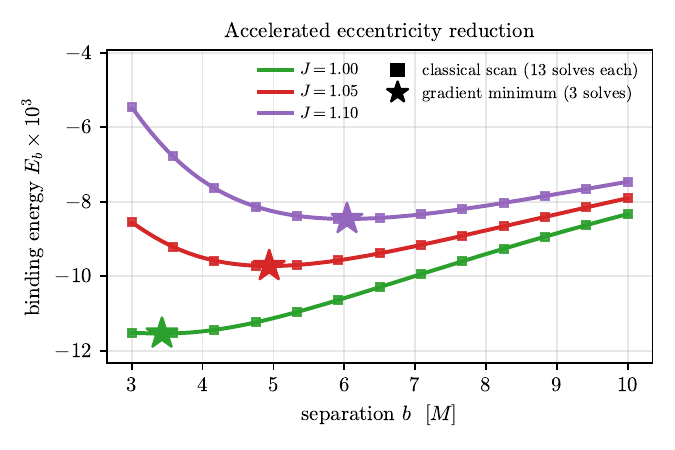}
  \caption{\label{fig:ecc}%
    \textbf{Accelerated eccentricity reduction.} At fixed angular momentum $J$ the
    binding energy $E_{\rm b}(b)$ has a minimum at the circular orbit; sweeping $J$ the
    minimum shifts outward, tracing the circular-orbit sequence. For each $J$ the
    differentiable $\partial E_{\rm b}/\partial b$ locates that minimum in three certified
    solves, against the classical certified-solve scan's thirteen (squares).
    In the shallow small-$J$ well the fixed scan mislocates the minimum, whereas the
    derivative still resolves it.}
\end{figure}

The classical method locates $b_{\rm circ}$ by \emph{scanning} $b$ at fixed $J$, i.e.\
one certified solve per point, and fitting the minimum. This workflow needs the
tangential momentum $P_{\rm t}$ as a \emph{free} axis (one scans $b$ at fixed
$J=2bP_{\rm t}$), which the quasi-circular model of Sec.~\ref{sec:model} deliberately
does not carry; we therefore build a dedicated small two-dimensional model in
$(b,P_{\rm t})$ (equal mass, no spin) by the identical machinery of
Secs.~\ref{sec:param}--\ref{sec:model}. On it the differentiable binding energy
exposes $\partial E_{\rm b}/\partial b|_J$ analytically, so a Newton root-find locates the
circular orbit on the interpolant and certifies at the end. Sweeping $J$
traces the circular-orbit sequence, whose minimum shifts outward with angular
momentum (Fig.~\ref{fig:ecc}). The gradient locates each $b_{\rm circ}$ in three
certified solves against the scan's thirteen, agreeing with it to
$6\times10^{-3}$ where the well is well resolved. It is moreover \emph{more
accurate} in the shallow, near-innermost-stable-orbit regime, where the fixed
thirteen-point scan mislocates the minimum by $0.05\,M$ in $b$ and the derivative still
resolves it: a black-box scan must sample ever more finely, at more certified
solves, exactly where the physics is most delicate. The same differentiable $E_{\rm b}$
reads the orbital eccentricity off the turning points, from $e=0$ at the circular
orbit to $e\approx0.42$ at the box edge, and so exposes
$\partial e/\partial\theta$ for gradient-based eccentricity reduction.

Both workflows make the same point: the differentiable, certified map turns a
black-box search (a control loop and an effective-potential scan) into a gradient
problem solved on the parametric model and closed out in a handful of certified
solves, where the black box instead pays a full certified elliptic solve at every trial.

\section{Discussion and conclusions}
\label{sec:conclusions}
We have introduced a differentiable parametric solver for binary-black-hole puncture initial data
that combines sparse interpolation with Newton refinement. The interpolation
supplies an inexpensive parameter-dependent initial guess, while the subsequent
elliptic solve certifies every returned datum to a prescribed constraint tolerance.
Because the construction is implemented in JAX, the resulting parameter-to-solution
map is also analytically differentiable with respect to the physical parameters. The
four-dimensional aligned-spin and eight-dimensional general-spin models demonstrate
that this construction remains practical across the aligned- and general-spin
parameter spaces of moving-puncture practice, while the parameter-targeting and
eccentricity-reduction examples show how its derivatives can reduce the number of
certified elliptic solves required in downstream workflows.

The two headline parametric models are genuinely three-dimensional and
quasi-circular, spanning the aligned- and general-spin families (Sec.~\ref{sec:model}).
What remains is scope. The most natural extension is
\emph{eccentricity}: relaxing the quasi-circularity condition to carry an
independent radial momentum adds an eccentricity axis on top of the orbiting family, by the same
machinery.

One limitation is inherited from the free data themselves, not from the
parametric method. Bowen--York, conformally flat data carry spurious (``junk'')
radiation that the early evolution must shed~\cite{Lovelace2009}, and cannot
represent near-extremal black holes: the \emph{horizon} spin of a Bowen--York
puncture saturates at $\chi\approx0.93$ no matter how large the free-data spin
parameter~\cite{CookYork1990,DainLoustoZlochower2008}, which is why
nearly-extremal-spin simulations use conformally curved (e.g.\ superposed
Kerr--Schild) initial data~\cite{Lovelace2008}. Within the moving-puncture
mainstream ($\chi\lesssim0.9$) the family built here is the standard input; and
because the parametric layer is solver-agnostic, a conformally curved solver could be swapped in
under the same interpolation, certification, and compression machinery.
We will explore this in future work.

Two accuracy caveats are worth stating explicitly. First, the spin-wall analysis
is quantitative only within the physical range ($\chi_{\max}\lesssim0.8$): pushing
the sampling interval toward and past extremal spin mixes the genuine (complex,
far) singularity with pre-asymptotic and spatial-floor effects, so the clean
single-$\chi_\ast$ characterization is reported for the physical regime, not
extrapolated beyond it. Second, the certified $\|R\|^{(v)}_\infty\le10^{-10}$ gate is
verified at the production grid ($44\times32\times8$); at higher meridian
resolution the whole-field equilibrated residual floors higher, because the stiff
$m^2/\rho^2$ term amplifies roundoff in the azimuthal modes that carry no source
content, while the physically populated modes stay three orders of magnitude
below. The floor is therefore set by empty modes rather than by physics, and
the certification tolerance should be read at the grid it is quoted for.

Finally, beyond the immediate use cases, the method supplies what a data-driven
model of binary-black-hole spacetimes needs at its input: a fast,
\emph{differentiable} initial-data generator across a physical parameter space
whose output satisfies the constraints to a checked tolerance. Such a generator can
serve as a training source and a warm-start oracle for those models, supplying
gradients $\dID$ for end-to-end parameter inference.

\begin{acknowledgments}
The authors would like to thank Miguel Bezares Figueroa, Thiago Assump\c{c}\~ao, Ignacio Brevis, and Lli\-bert Arest\'e Sal\'o for encouraging and interesting discussions.
This work was developed with the assistance of Claude Code (Anthropic).
This research was supported by the Research Foundation -- Flanders (FWO; Grant No.\ I000725N and I002123N).
\end{acknowledgments}

\section*{Data and code availability}
The whole implementation is publicly available at
\url{https://github.com/LemaitreModels/LMID-conformally-flat-puncture}.

\appendix

\section{Validation against TwoPunctures}
\label{sec:validation}
We validate the initial-data solver against the Two\-Punctures
code~\cite{AnsorgBruegmannTichy2004}, using its standalone C port generated by
the NRPy suite~\cite{RuchlinEtienneBaumgarte2018,NRPy} and compiled against
GSL~\cite{GSL}. We compare the quasi-circular data over the
production box (App.~\ref{sec:validation:qc}) and, through an independent
evolution code's constraint diagnostic, on an evolution grid
(App.~\ref{sec:validation:constraints}). In the
axi\-symmetric limit the two codes agree to near machine precision.

\begin{figure}[t]
  \centering
  \includegraphics[width=\columnwidth]{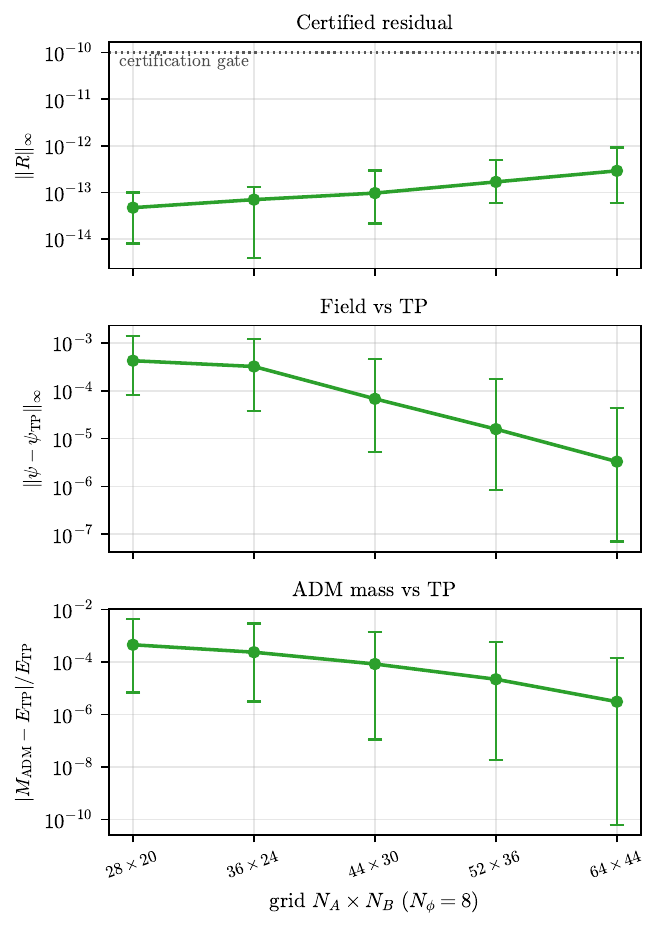}
  \caption{\label{fig:tpval}%
    \textbf{Quasi-circular data validated against Two\-Punctures, across the
    production box.} Each point is a median over $100$ quasi-circular
    configurations drawn from the production box by a Latin hypercube, with
    whiskers spanning the sample minimum to maximum; the shared resolution axis
    is the meridional sequence at fixed $N_\phi=8$, and every configuration is
    compared with one TwoPunctures solve at $72\times72\times12$.
    The panels are, from top to bottom, the certified residual; the pointwise field difference; and
    the ADM mass comparison. The field difference falls with resolution while the residual rises: the rise is
    roundoff amplification in unpopulated high-$m$ modes, not a loss of convergence, and every
    configuration stays below the certification gate $10^{-10}$ (dotted), in the
    row-equilibrated norm $\|R\|^{(v)}_\infty$, at
    every resolution.}
\end{figure}

\subsection{Quasi-circular data against TwoPunctures}
\label{sec:validation:qc}

We validate the quasi-circular configurations used in Sec.~\ref{sec:model}.
Because our prolate coordinate system places the punctures on the $z$ axis,
whereas the standard Two\-Punctures convention places them on the $x$ axis, we
first rotate both data sets to a common Cartesian frame. We verified the
mapping between the bare masses, half-separation, and Bowen--York momenta and
spins used by the two codes; in the equal-mass nonspinning limit the
post-Newtonian expression of Sec.~\ref{sec:model} approaches the Newtonian
circular-orbit result, which fixes the separation and momentum conventions.
Both codes employ the decomposition $\psi=\psi_{\rm BL}+u$, so the conformal
factor is compared directly at common physical points.

So that the comparison does not rest on one arbitrary parameter point, we draw
$100$ configurations from the production box by a Latin hypercube and walk the
same meridional resolution sequence for each, at fixed $N_\phi=8$. Every
configuration is compared with a single TwoPunctures solve at
$72\times72\times12$, which out-resolves the finest of our own grids in every
direction. Ten further configurations placed deliberately at the box edges are
reported separately, so that they do not bias the sample statistics.

The pointwise field difference converges across the sampled box, as shown in
Fig.~\ref{fig:tpval}.\footnote{Pointwise differences are estimated on a fixed
set of $70$ interior probe points, placed clear of both punctures and covering
the full azimuthal range. The reference solve is converged well below every difference reported:
its field moves by below $10^{-11}$ across the sampled box between its own
$72\times72$ and $96\times96$ meridional resolutions, and its azimuthal truncation
contributes at most $2\times10^{-8}$. The comparison therefore reflects our discretization
rather than the reference.} The minimum,
median and maximum of the sample each fall monotonically along the sequence,
although $28$ of the $100$ configurations show one non-monotone step somewhere
along the ladder, as a supremum norm of a spectral error may; we therefore claim
convergence of the distribution rather than of every member of it. What limits
the disagreement is the meridional resolution rather than the nonlinear solve:
raising $N_\phi$ alongside $(N_A,N_B)$ leaves the field difference unchanged to
three digits at every resolution.

The certified residual behaves in the opposite direction
(Fig.~\ref{fig:tpval}). Its median rises along the same
sequence, while remaining below the certification tolerance $10^{-10}$ in $\|R\|^{(v)}_\infty$ for every one of the $100$ configurations
at every resolution. That the field difference falls while the residual grows
identifies the rise as roundoff amplification in unpopulated high-$m$ modes
rather than a loss of convergence. The direction of refinement decides the size
of the effect: raising $N_\phi$ instead makes the residual rise by six orders of
magnitude and breach the tolerance, which is why the production sequence refines
the meridian.

The integral (ADM mass) comparison, shown in Fig.~\ref{fig:tpval}, converges faster than
the pointwise (field) one. However, the two remain comparable.

\begin{figure}[t]
  \centering
  \includegraphics[width=\columnwidth]{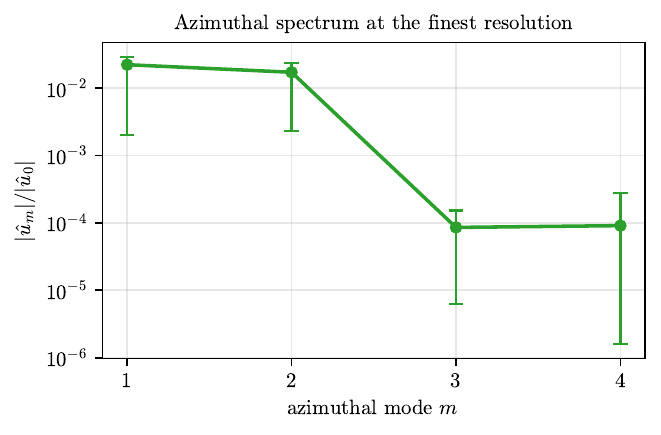}
  \caption{\label{fig:tpspec}%
    \textbf{Azimuthal structure of the quasi-circular data.} Azimuthal Fourier
    spectrum $|\hat u_m|/|\hat u_0|$ at the finest resolution: median with
    min--max whiskers over the same $100$ configurations as Fig.~\ref{fig:tpval},
    with $m=0$ the normalization and not shown. The
    last point is that grid's Nyquist mode (dotted): $\sin 4\phi$ vanishes
    identically on $N_\phi$ equispaced nodes, so it carries only the
    $\cos 4\phi$ component, one real degree of freedom against two for
    $m=1$--$3$, and is not comparable like-for-like with them.}
\end{figure}

The azimuthal Fourier spectrum in Fig.~\ref{fig:tpspec} shows where the
non-axisymmetry sits. These data are non-axisymmetric by construction: the
tangential momentum populates $m=2$ and generic spins populate $m=1$. Across the
sample both sit at the percent level. The spectrum then falls by more than two decades; $m=4$ is the
Nyquist amplitude of the $N_\phi=8$ grid and bounds rather than resolves the
high-$m$ content. A small number of Fourier modes accordingly resolves these
data, as the refinement above establishes independently.

Two checks that the quasi-circular family cannot supply, because it is never
axisymmetric, complete the picture. First, for four head-on configurations with
both spins along the collision axis, $(b,\,q,\,\chi^{A}_{z}=\chi^{B}_{z}) =
(4,1,0)$, $(4,1,0.5)$, $(4,3,0.5)$ and $(3,3,0.9)$,
all $m\ge1$ modes remain below $10^{-16}$, confirming that the Fourier-in-$\phi$
operator does not manufacture spurious non-axisymmetry. Second, in this
axisymmetric limit the pointwise field agreement approaches machine precision:
for a non-spinning equal-mass configuration with $b=3$ and axial momentum
$P=0.5$, solved on
the same grid sequence and compared against the same reference, the two
independently discretized spectral codes agree in $\psi$ to $2.5\times10^{-10}$
in the supremum norm, and in the
total ADM mass to a relative difference of $1.2\times10^{-11}$; the certified
residual there is $9.3\times10^{-15}$. With no azimuthal structure to limit
either code, this is the most stringent code-to-code reference available for the
solver.

\subsection{Constraint violation on an evolution grid}
\label{sec:validation:constraints}

\begin{figure}[t]
  \centering
  \includegraphics[width=\columnwidth]{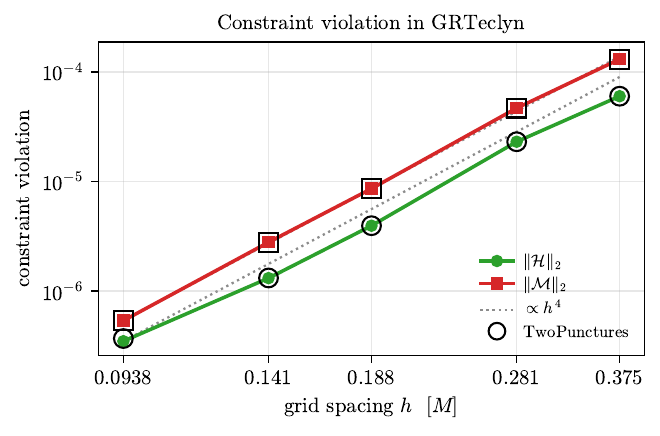}
  \caption{\label{fig:constraints}%
    \textbf{Constraint violation of the initial data, as measured by GRTeclyn.} Volume-normalized bulk $L_2$ norms of the Hamiltonian and
    momentum constraints at $t=0$, computed by GRTeclyn~\cite{GRTeclyn} with its
    own fourth-order stencils,
    against the grid spacing $h$ on a uniform ladder
    ($N^3$ with $N=48$, $64$, $96$, $128$, $192$) across a box of half-width
    $9\,M$, with the punctures excluded to radius $1.5\,M$.
    The configuration is the axisymmetric anchor of this appendix: equal bare
    masses $m_A=m_B=0.5$, zero spins, half-separation $b=3$ and axial momentum
    $P=0.5$.
    Black open markers repeat the
    measurement with the conformal factor supplied by Two\-Punctures.}
\end{figure}

As a final validation, we hand the initial data to GR\-Teclyn~\cite{GRTeclyn}, an
independent numerical-relativity code, and let it measure the
constraint violation with its own machinery. GRTeclyn evaluates the Hamiltonian
and momentum constraints with fourth-order stencils, working from the full evolution
variables rather than from the conformal factor directly, so
the measurement also exercises the conversion into evolution form that a
simulation actually performs.

The data are transferred as the solved conformal factor on its spectral grid and
evaluated in the evolution code by a direct port of the interpolation used here.
Since the transfer accepts a conformal factor from any source, we also run the TwoPunctures solution through
it, on the same spectral grid, so that the initial data is the only difference
between the two measurements. Figure~\ref{fig:constraints} shows both norms
falling at fourth order with the TwoPunctures points lying on our curves:
one cannot distinguish the two solutions until $h=0.094\,M$, and
there only at the level of $6\%$ of a violation of $3.5\times10^{-7}$. The two
conformal factors themselves agree to $3.6\times10^{-10}$ relative.

\section{Separable form of the per-mode block}
\label{sec:separable}

This appendix records the reduction quoted in Sec.~\ref{sec:solve:newton}. Let
$\mathcal{P}\equiv(1-A^2)^2/(b^2\mathcal{D})$ denote the common prefactor of
Eq.~\eqref{eq:abt_laplacian}, and let
$W\equiv\mathrm{diag}\big[(1-B_j^2)^{|m|/2}\big]$ be the associated-Legendre
factor of Eq.~\eqref{eq:basis}. The mode-$m$ unknown is carried as the smooth
factored array
\begin{equation}
  V^{(m)} \ = \ W^{-1}\,\widehat U^{(m)},
  \label{eq:smooth_mode}
\end{equation}
of shape $(N_A{+}1)\times N_B$, with $A$ indexing rows and $B$ columns.

Every coefficient of Eq.~\eqref{eq:abt_laplacian} is $\mathcal{P}$ times a
function of $A$ alone or of $B$ alone: the two derivative terms by inspection, and
the centrifugal term by Eq.~\eqref{eq:centrifugal_split}. Dividing by
$\mathcal{P}$ therefore leaves no $A$--$B$ mixing. On the interior rows, i.e.\ the
only rows where $\mathcal{P}^{-1}$ is defined, the $A=1$ boundary rows having been
row-replaced, the block factors as
\begin{equation}
  \mathrm{diag}\big(\mathcal{P}^{-1}\big)\,M_0^{(m)}
  \, = \, \big(I\otimes W\big)\,
        \big(L_A^{(m)}\otimes I \, + \, I\otimes L_B^{(m)}\big),
  \label{eq:kronecker_sum}
\end{equation}
a diagonal factor times a true Kronecker sum of two one-dimensional operators,
$L_A^{(m)}$ of size $N_A{+}1$ and $L_B^{(m)}$ of size $N_B$. The two are not
symmetric in construction: $L_A^{(m)}$ is the bare $A$-operator, whereas
$L_B^{(m)}$ acts on the smooth factor of Eq.~\eqref{eq:smooth_mode} and so carries
the analytic first and second derivatives of $W$ along with $W$ itself.

The bracket in Eq.~\eqref{eq:kronecker_sum} acts on $V^{(m)}$ as the Sylvester
operator $L_A^{(m)}V^{(m)}+V^{(m)} (L_B^{(m)})^{T}$. Diagonalizing each
one-dimensional factor once, $L_A^{(m)}=S_A\Lambda_AS_A^{-1}$ and
$\big(L_B^{(m)}\big)^{\!T}=S_B\Lambda_BS_B^{-1}$, reduces the solve to a division
by $\lambda^A_i+\lambda^B_j$ in the transformed variables, the right-hand side
having first been divided by $W$. One apply then costs
$O\big(N_AN_B(N_A{+}N_B)\big)$ after an $O(N_A^3+N_B^3)$ setup. Both eigenproblems
are nonsymmetric, and the conditioning of $S_A$ and $S_B$ is what limits the
attainable accuracy of the inverse.

The boundary rows preserve this form because they are one-dimensional. The
$A$-direction conditions of Sec.~\ref{sec:solve:abt} (Dirichlet at $A=1$, and at
$A=0$ either the $m=0$ Neumann row or $u_m=0$) involve $A$ alone, so they are
imposed inside $L_A^{(m)}$ and then eliminated exactly: writing $\partial$ for
those two rows and $i$ for the interior, solving the boundary equations for
$V_\partial$ and substituting into the interior rows replaces $L_A^{(m)}$ by the
Schur complement $L_{ii}-L_{i\partial}L_{\partial\partial}^{-1}L_{\partial i}$
(static condensation), again a one-dimensional
operator in $A$. A condition coupling $A$ and $B$ would not permit this, and the
Kronecker structure would be lost.

Every interior coefficient finally carries exactly one factor $b^{-2}$, through
$\mathcal{P}$ and through the centrifugal term alike, while the boundary rows are
$b$-free. Hence $M_0^{(m)}(b)=D(b)\,M_0^{(m)}(1)$ with $D$ diagonal, a left scaling
that row equilibration cancels identically. The one-dimensional factors are
therefore built once per grid, and the separation enters only as a scalar row
factor at apply time.

\bibliographystyle{apsrev4-2}
\bibliography{references}

\begin{thebibliography}{41}%
\makeatletter
\providecommand \@ifxundefined [1]{%
 \@ifx{#1\undefined}
}%
\providecommand \@ifnum [1]{%
 \ifnum #1\expandafter \@firstoftwo
 \else \expandafter \@secondoftwo
 \fi
}%
\providecommand \@ifx [1]{%
 \ifx #1\expandafter \@firstoftwo
 \else \expandafter \@secondoftwo
 \fi
}%
\providecommand \natexlab [1]{#1}%
\providecommand \enquote  [1]{``#1''}%
\providecommand \bibnamefont  [1]{#1}%
\providecommand \bibfnamefont [1]{#1}%
\providecommand \citenamefont [1]{#1}%
\providecommand \href@noop [0]{\@secondoftwo}%
\providecommand \href [0]{\begingroup \@sanitize@url \@href}%
\providecommand \@href[1]{\@@startlink{#1}\@@href}%
\providecommand \@@href[1]{\endgroup#1\@@endlink}%
\providecommand \@sanitize@url [0]{\catcode `\\12\catcode `\$12\catcode
  `\&12\catcode `\#12\catcode `\^12\catcode `\_12\catcode `\%12\relax}%
\providecommand \@@startlink[1]{}%
\providecommand \@@endlink[0]{}%
\providecommand \url  [0]{\begingroup\@sanitize@url \@url }%
\providecommand \@url [1]{\endgroup\@href {#1}{\urlprefix }}%
\providecommand \urlprefix  [0]{URL }%
\providecommand \Eprint [0]{\href }%
\providecommand \doibase [0]{https://doi.org/}%
\providecommand \selectlanguage [0]{\@gobble}%
\providecommand \bibinfo  [0]{\@secondoftwo}%
\providecommand \bibfield  [0]{\@secondoftwo}%
\providecommand \translation [1]{[#1]}%
\providecommand \BibitemOpen [0]{}%
\providecommand \bibitemStop [0]{}%
\providecommand \bibitemNoStop [0]{.\EOS\space}%
\providecommand \EOS [0]{\spacefactor3000\relax}%
\providecommand \BibitemShut  [1]{\csname bibitem#1\endcsname}%
\let\auto@bib@innerbib\@empty
\bibitem [{\citenamefont {York}(1979)}]{York1979}%
  \BibitemOpen
  \bibfield  {author} {\bibinfo {author} {\bibfnamefont {J.~W.}\ \bibnamefont
  {York}, \bibfnamefont {Jr.}},\ }in\ \href@noop {} {\emph {\bibinfo
  {booktitle} {Sources of Gravitational Radiation}}},\ \bibinfo {editor}
  {edited by\ \bibinfo {editor} {\bibfnamefont {L.~L.}\ \bibnamefont {Smarr}}}\
  (\bibinfo  {publisher} {Cambridge University Press},\ \bibinfo {address}
  {Cambridge, England},\ \bibinfo {year} {1979})\ pp.\ \bibinfo {pages}
  {83--126}\BibitemShut {NoStop}%
\bibitem [{\citenamefont {Bowen}\ and\ \citenamefont
  {York}(1980)}]{BowenYork1980}%
  \BibitemOpen
  \bibfield  {author} {\bibinfo {author} {\bibfnamefont {J.~M.}\ \bibnamefont
  {Bowen}}\ and\ \bibinfo {author} {\bibfnamefont {J.~W.}\ \bibnamefont {York},
  \bibfnamefont {Jr.}},\ }\href {https://doi.org/10.1103/PhysRevD.21.2047}
  {\bibfield  {journal} {\bibinfo  {journal} {Phys. Rev. D}\ }\textbf {\bibinfo
  {volume} {21}},\ \bibinfo {pages} {2047} (\bibinfo {year}
  {1980})}\BibitemShut {NoStop}%
\bibitem [{\citenamefont {Brandt}\ and\ \citenamefont
  {Br{\"u}gmann}(1997)}]{BrandtBruegmann1997}%
  \BibitemOpen
  \bibfield  {author} {\bibinfo {author} {\bibfnamefont {S.}~\bibnamefont
  {Brandt}}\ and\ \bibinfo {author} {\bibfnamefont {B.}~\bibnamefont
  {Br{\"u}gmann}},\ }\href {https://doi.org/10.1103/PhysRevLett.78.3606}
  {\bibfield  {journal} {\bibinfo  {journal} {Phys. Rev. Lett.}\ }\textbf
  {\bibinfo {volume} {78}},\ \bibinfo {pages} {3606} (\bibinfo {year}
  {1997})}\BibitemShut {NoStop}%
\bibitem [{\citenamefont {Cook}(2000)}]{Cook2000}%
  \BibitemOpen
  \bibfield  {author} {\bibinfo {author} {\bibfnamefont {G.~B.}\ \bibnamefont
  {Cook}},\ }\href {https://doi.org/10.12942/lrr-2000-5} {\bibfield  {journal}
  {\bibinfo  {journal} {Living Rev. Relativ.}\ }\textbf {\bibinfo {volume}
  {3}},\ \bibinfo {pages} {5} (\bibinfo {year} {2000})}\BibitemShut {NoStop}%
\bibitem [{\citenamefont {Mendes}\ \emph {et~al.}(2025)\citenamefont {Mendes},
  \citenamefont {Vu}, \citenamefont {Long}, \citenamefont {Pfeiffer},\ and\
  \citenamefont {Owen}}]{Mendes2025}%
  \BibitemOpen
  \bibfield  {author} {\bibinfo {author} {\bibfnamefont {I.~B.}\ \bibnamefont
  {Mendes}}, \bibinfo {author} {\bibfnamefont {N.~L.}\ \bibnamefont {Vu}},
  \bibinfo {author} {\bibfnamefont {O.}~\bibnamefont {Long}}, \bibinfo {author}
  {\bibfnamefont {H.~P.}\ \bibnamefont {Pfeiffer}},\ and\ \bibinfo {author}
  {\bibfnamefont {R.}~\bibnamefont {Owen}},\ }\href
  {https://doi.org/10.1103/zh31-bbtm} {\bibfield  {journal} {\bibinfo
  {journal} {Phys. Rev. D}\ }\textbf {\bibinfo {volume} {112}},\ \bibinfo
  {pages} {124049} (\bibinfo {year} {2025})}\BibitemShut {NoStop}%
\bibitem [{\citenamefont {Field}\ \emph {et~al.}(2011)\citenamefont {Field},
  \citenamefont {Galley}, \citenamefont {Herrmann}, \citenamefont {Hesthaven},
  \citenamefont {Ochsner},\ and\ \citenamefont {Tiglio}}]{FieldGalley2011}%
  \BibitemOpen
  \bibfield  {author} {\bibinfo {author} {\bibfnamefont {S.~E.}\ \bibnamefont
  {Field}}, \bibinfo {author} {\bibfnamefont {C.~R.}\ \bibnamefont {Galley}},
  \bibinfo {author} {\bibfnamefont {F.}~\bibnamefont {Herrmann}}, \bibinfo
  {author} {\bibfnamefont {J.~S.}\ \bibnamefont {Hesthaven}}, \bibinfo {author}
  {\bibfnamefont {E.}~\bibnamefont {Ochsner}},\ and\ \bibinfo {author}
  {\bibfnamefont {M.}~\bibnamefont {Tiglio}},\ }\href
  {https://doi.org/10.1103/PhysRevLett.106.221102} {\bibfield  {journal}
  {\bibinfo  {journal} {Phys. Rev. Lett.}\ }\textbf {\bibinfo {volume} {106}},\
  \bibinfo {pages} {221102} (\bibinfo {year} {2011})}\BibitemShut {NoStop}%
\bibitem [{\citenamefont {Field}\ \emph {et~al.}(2014)\citenamefont {Field},
  \citenamefont {Galley}, \citenamefont {Hesthaven}, \citenamefont {Kaye},\
  and\ \citenamefont {Tiglio}}]{Field2014}%
  \BibitemOpen
  \bibfield  {author} {\bibinfo {author} {\bibfnamefont {S.~E.}\ \bibnamefont
  {Field}}, \bibinfo {author} {\bibfnamefont {C.~R.}\ \bibnamefont {Galley}},
  \bibinfo {author} {\bibfnamefont {J.~S.}\ \bibnamefont {Hesthaven}}, \bibinfo
  {author} {\bibfnamefont {J.}~\bibnamefont {Kaye}},\ and\ \bibinfo {author}
  {\bibfnamefont {M.}~\bibnamefont {Tiglio}},\ }\href
  {https://doi.org/10.1103/PhysRevX.4.031006} {\bibfield  {journal} {\bibinfo
  {journal} {Phys. Rev. X}\ }\textbf {\bibinfo {volume} {4}},\ \bibinfo {pages}
  {031006} (\bibinfo {year} {2014})}\BibitemShut {NoStop}%
\bibitem [{\citenamefont {Blackman}\ \emph {et~al.}(2015)\citenamefont
  {Blackman}, \citenamefont {Field}, \citenamefont {Galley}, \citenamefont
  {Szil{\'a}gyi}, \citenamefont {Scheel}, \citenamefont {Tiglio},\ and\
  \citenamefont {Hemberger}}]{Blackman2015}%
  \BibitemOpen
  \bibfield  {author} {\bibinfo {author} {\bibfnamefont {J.}~\bibnamefont
  {Blackman}}, \bibinfo {author} {\bibfnamefont {S.~E.}\ \bibnamefont {Field}},
  \bibinfo {author} {\bibfnamefont {C.~R.}\ \bibnamefont {Galley}}, \bibinfo
  {author} {\bibfnamefont {B.}~\bibnamefont {Szil{\'a}gyi}}, \bibinfo {author}
  {\bibfnamefont {M.~A.}\ \bibnamefont {Scheel}}, \bibinfo {author}
  {\bibfnamefont {M.}~\bibnamefont {Tiglio}},\ and\ \bibinfo {author}
  {\bibfnamefont {D.~A.}\ \bibnamefont {Hemberger}},\ }\href
  {https://doi.org/10.1103/PhysRevLett.115.121102} {\bibfield  {journal}
  {\bibinfo  {journal} {Phys. Rev. Lett.}\ }\textbf {\bibinfo {volume} {115}},\
  \bibinfo {pages} {121102} (\bibinfo {year} {2015})}\BibitemShut {NoStop}%
\bibitem [{\citenamefont {Varma}\ \emph {et~al.}(2019)\citenamefont {Varma},
  \citenamefont {Field}, \citenamefont {Scheel}, \citenamefont {Blackman},
  \citenamefont {Gerosa}, \citenamefont {Stein}, \citenamefont {Kidder},\ and\
  \citenamefont {Pfeiffer}}]{Varma2019}%
  \BibitemOpen
  \bibfield  {author} {\bibinfo {author} {\bibfnamefont {V.}~\bibnamefont
  {Varma}}, \bibinfo {author} {\bibfnamefont {S.~E.}\ \bibnamefont {Field}},
  \bibinfo {author} {\bibfnamefont {M.~A.}\ \bibnamefont {Scheel}}, \bibinfo
  {author} {\bibfnamefont {J.}~\bibnamefont {Blackman}}, \bibinfo {author}
  {\bibfnamefont {D.}~\bibnamefont {Gerosa}}, \bibinfo {author} {\bibfnamefont
  {L.~C.}\ \bibnamefont {Stein}}, \bibinfo {author} {\bibfnamefont {L.~E.}\
  \bibnamefont {Kidder}},\ and\ \bibinfo {author} {\bibfnamefont {H.~P.}\
  \bibnamefont {Pfeiffer}},\ }\href
  {https://doi.org/10.1103/PhysRevResearch.1.033015} {\bibfield  {journal}
  {\bibinfo  {journal} {Phys. Rev. Research}\ }\textbf {\bibinfo {volume}
  {1}},\ \bibinfo {pages} {033015} (\bibinfo {year} {2019})}\BibitemShut
  {NoStop}%
\bibitem [{\citenamefont {Reed}\ \emph {et~al.}(2024)\citenamefont {Reed},
  \citenamefont {Somasundaram}, \citenamefont {De}, \citenamefont {Armstrong},
  \citenamefont {Giuliani}, \citenamefont {Capano}, \citenamefont {Brown},\
  and\ \citenamefont {Tews}}]{Reed2024}%
  \BibitemOpen
  \bibfield  {author} {\bibinfo {author} {\bibfnamefont {B.~T.}\ \bibnamefont
  {Reed}}, \bibinfo {author} {\bibfnamefont {R.}~\bibnamefont {Somasundaram}},
  \bibinfo {author} {\bibfnamefont {S.}~\bibnamefont {De}}, \bibinfo {author}
  {\bibfnamefont {C.~L.}\ \bibnamefont {Armstrong}}, \bibinfo {author}
  {\bibfnamefont {P.}~\bibnamefont {Giuliani}}, \bibinfo {author}
  {\bibfnamefont {C.}~\bibnamefont {Capano}}, \bibinfo {author} {\bibfnamefont
  {D.~A.}\ \bibnamefont {Brown}},\ and\ \bibinfo {author} {\bibfnamefont
  {I.}~\bibnamefont {Tews}},\ }\href {https://doi.org/10.3847/1538-4357/ad737c}
  {\bibfield  {journal} {\bibinfo  {journal} {Astrophys. J.}\ }\textbf
  {\bibinfo {volume} {974}},\ \bibinfo {pages} {285} (\bibinfo {year}
  {2024})},\ \Eprint {https://arxiv.org/abs/2405.20558} {arXiv:2405.20558
  [astro-ph.HE]} \BibitemShut {NoStop}%
\bibitem [{\citenamefont {Babu{\v s}ka}\ \emph {et~al.}(2007)\citenamefont
  {Babu{\v s}ka}, \citenamefont {Nobile},\ and\ \citenamefont
  {Tempone}}]{BabuskaNobileTempone2007}%
  \BibitemOpen
  \bibfield  {author} {\bibinfo {author} {\bibfnamefont {I.}~\bibnamefont
  {Babu{\v s}ka}}, \bibinfo {author} {\bibfnamefont {F.}~\bibnamefont
  {Nobile}},\ and\ \bibinfo {author} {\bibfnamefont {R.}~\bibnamefont
  {Tempone}},\ }\href {https://doi.org/10.1137/050645142} {\bibfield  {journal}
  {\bibinfo  {journal} {SIAM J. Numer. Anal.}\ }\textbf {\bibinfo {volume}
  {45}},\ \bibinfo {pages} {1005} (\bibinfo {year} {2007})}\BibitemShut
  {NoStop}%
\bibitem [{\citenamefont {Xiu}\ and\ \citenamefont
  {Hesthaven}(2005)}]{XiuHesthaven2005}%
  \BibitemOpen
  \bibfield  {author} {\bibinfo {author} {\bibfnamefont {D.}~\bibnamefont
  {Xiu}}\ and\ \bibinfo {author} {\bibfnamefont {J.~S.}\ \bibnamefont
  {Hesthaven}},\ }\href {https://doi.org/10.1137/040615201} {\bibfield
  {journal} {\bibinfo  {journal} {SIAM J. Sci. Comput.}\ }\textbf {\bibinfo
  {volume} {27}},\ \bibinfo {pages} {1118} (\bibinfo {year}
  {2005})}\BibitemShut {NoStop}%
\bibitem [{\citenamefont {Nobile}\ \emph {et~al.}(2008)\citenamefont {Nobile},
  \citenamefont {Tempone},\ and\ \citenamefont
  {Webster}}]{NobileTemponeWebster2008}%
  \BibitemOpen
  \bibfield  {author} {\bibinfo {author} {\bibfnamefont {F.}~\bibnamefont
  {Nobile}}, \bibinfo {author} {\bibfnamefont {R.}~\bibnamefont {Tempone}},\
  and\ \bibinfo {author} {\bibfnamefont {C.~G.}\ \bibnamefont {Webster}},\
  }\href {https://doi.org/10.1137/060663660} {\bibfield  {journal} {\bibinfo
  {journal} {SIAM J. Numer. Anal.}\ }\textbf {\bibinfo {volume} {46}},\
  \bibinfo {pages} {2309} (\bibinfo {year} {2008})}\BibitemShut {NoStop}%
\bibitem [{\citenamefont {Haasdonk}(2017)}]{Haasdonk2017}%
  \BibitemOpen
  \bibfield  {author} {\bibinfo {author} {\bibfnamefont {B.}~\bibnamefont
  {Haasdonk}},\ }in\ \href {https://doi.org/10.1137/1.9781611974829.ch2} {\emph
  {\bibinfo {booktitle} {Model Reduction and Approximation: Theory and
  Algorithms}}},\ \bibinfo {series and number} {\bibinfo {series}
  {Computational Science and Engineering}\ No.~\bibinfo {number} {15}},\
  \bibinfo {editor} {edited by\ \bibinfo {editor} {\bibfnamefont
  {P.}~\bibnamefont {Benner}}, \bibinfo {editor} {\bibfnamefont
  {A.}~\bibnamefont {Cohen}}, \bibinfo {editor} {\bibfnamefont
  {M.}~\bibnamefont {Ohlberger}},\ and\ \bibinfo {editor} {\bibfnamefont
  {K.}~\bibnamefont {Willcox}}}\ (\bibinfo  {publisher} {SIAM},\ \bibinfo
  {address} {Philadelphia, PA},\ \bibinfo {year} {2017})\ Chap.~\bibinfo
  {chapter} {2}, pp.\ \bibinfo {pages} {65--136}\BibitemShut {NoStop}%
\bibitem [{\citenamefont {Hesthaven}\ \emph {et~al.}(2016)\citenamefont
  {Hesthaven}, \citenamefont {Rozza},\ and\ \citenamefont
  {Stamm}}]{HesthavenRozzaStamm2016}%
  \BibitemOpen
  \bibfield  {author} {\bibinfo {author} {\bibfnamefont {J.~S.}\ \bibnamefont
  {Hesthaven}}, \bibinfo {author} {\bibfnamefont {G.}~\bibnamefont {Rozza}},\
  and\ \bibinfo {author} {\bibfnamefont {B.}~\bibnamefont {Stamm}},\ }\href
  {https://doi.org/10.1007/978-3-319-22470-1} {\emph {\bibinfo {title}
  {Certified Reduced Basis Methods for Parametrized Partial Differential
  Equations}}},\ SpringerBriefs in Mathematics\ (\bibinfo  {publisher}
  {Springer International Publishing},\ \bibinfo {address} {Cham},\ \bibinfo
  {year} {2016})\BibitemShut {NoStop}%
\bibitem [{\citenamefont {Trefethen}(2019)}]{Trefethen2019}%
  \BibitemOpen
  \bibfield  {author} {\bibinfo {author} {\bibfnamefont {L.~N.}\ \bibnamefont
  {Trefethen}},\ }\href {https://doi.org/10.1137/1.9781611975949} {\emph
  {\bibinfo {title} {Approximation Theory and Approximation Practice, Extended
  Edition}}}\ (\bibinfo  {publisher} {Society for Industrial and Applied
  Mathematics},\ \bibinfo {address} {Philadelphia, PA},\ \bibinfo {year}
  {2019})\BibitemShut {NoStop}%
\bibitem [{\citenamefont {Tommasini}\ \emph {et~al.}(2026)\citenamefont
  {Tommasini}, \citenamefont {Vu}, \citenamefont {Scheel},\ and\ \citenamefont
  {Teukolsky}}]{Tommasini2026}%
  \BibitemOpen
  \bibfield  {author} {\bibinfo {author} {\bibfnamefont {V.}~\bibnamefont
  {Tommasini}}, \bibinfo {author} {\bibfnamefont {N.~L.}\ \bibnamefont {Vu}},
  \bibinfo {author} {\bibfnamefont {M.~A.}\ \bibnamefont {Scheel}},\ and\
  \bibinfo {author} {\bibfnamefont {S.~A.}\ \bibnamefont {Teukolsky}},\
  }\href@noop {} {\bibinfo {title} {Data-driven acceleration of eccentricity
  reduction for binary black hole simulations}} (\bibinfo {year} {2026}),\
  \Eprint {https://arxiv.org/abs/2604.22021} {arXiv:2604.22021 [gr-qc]}
  \BibitemShut {NoStop}%
\bibitem [{\citenamefont {Zhou}\ \emph {et~al.}(2026)\citenamefont {Zhou},
  \citenamefont {Ma}, \citenamefont {Cao}, \citenamefont {Wu}, \citenamefont
  {Jin}, \citenamefont {Feng}, \citenamefont {Huang}, \citenamefont {Zhao},\
  and\ \citenamefont {Wu}}]{Zhou2026}%
  \BibitemOpen
  \bibfield  {author} {\bibinfo {author} {\bibfnamefont {Y.-C.}\ \bibnamefont
  {Zhou}}, \bibinfo {author} {\bibfnamefont {H.}~\bibnamefont {Ma}}, \bibinfo
  {author} {\bibfnamefont {Z.}~\bibnamefont {Cao}}, \bibinfo {author}
  {\bibfnamefont {T.}~\bibnamefont {Wu}}, \bibinfo {author} {\bibfnamefont
  {H.-B.}\ \bibnamefont {Jin}}, \bibinfo {author} {\bibfnamefont
  {X.}~\bibnamefont {Feng}}, \bibinfo {author} {\bibfnamefont {S.}~\bibnamefont
  {Huang}}, \bibinfo {author} {\bibfnamefont {Z.-C.}\ \bibnamefont {Zhao}},\
  and\ \bibinfo {author} {\bibfnamefont {Y.-L.}\ \bibnamefont {Wu}},\
  }\href@noop {} {\bibinfo {title} {Solving {Hamiltonian} constraint equation
  with physics-informed neural networks}} (\bibinfo {year} {2026}),\ \bibinfo
  {note} {accepted for publication in Phys. Rev. D},\ \Eprint
  {https://arxiv.org/abs/2607.06002} {arXiv:2607.06002 [gr-qc]} \BibitemShut
  {NoStop}%
\bibitem [{\citenamefont {Ogurol}\ \emph {et~al.}(2025)\citenamefont {Ogurol},
  \citenamefont {Boybeyi},\ and\ \citenamefont {Tekin}}]{Ogurol2025}%
  \BibitemOpen
  \bibfield  {author} {\bibinfo {author} {\bibfnamefont {L.}~\bibnamefont
  {Ogurol}}, \bibinfo {author} {\bibfnamefont {T.}~\bibnamefont {Boybeyi}},\
  and\ \bibinfo {author} {\bibfnamefont {B.}~\bibnamefont {Tekin}},\
  }\href@noop {} {\bibinfo {title} {Analytical perturbative construction of
  initial data for binary black holes up to third order in spin and momentum}}
  (\bibinfo {year} {2025}),\ \Eprint {https://arxiv.org/abs/2509.11144}
  {arXiv:2509.11144 [gr-qc]} \BibitemShut {NoStop}%
\bibitem [{\citenamefont {Cranganore}\ \emph {et~al.}(2025)\citenamefont
  {Cranganore}, \citenamefont {Bodnar}, \citenamefont {Berzins},\ and\
  \citenamefont {Brandstetter}}]{Cranganore2025}%
  \BibitemOpen
  \bibfield  {author} {\bibinfo {author} {\bibfnamefont {S.~S.}\ \bibnamefont
  {Cranganore}}, \bibinfo {author} {\bibfnamefont {A.}~\bibnamefont {Bodnar}},
  \bibinfo {author} {\bibfnamefont {A.}~\bibnamefont {Berzins}},\ and\ \bibinfo
  {author} {\bibfnamefont {J.}~\bibnamefont {Brandstetter}},\ }\href@noop {}
  {\bibinfo {title} {Einstein fields: A neural perspective to computational
  general relativity}} (\bibinfo {year} {2025}),\ \bibinfo {note} {accepted at
  ICLR 2026},\ \Eprint {https://arxiv.org/abs/2507.11589} {arXiv:2507.11589
  [cs.LG]} \BibitemShut {NoStop}%
\bibitem [{\citenamefont {Bradbury}\ \emph {et~al.}(2018)\citenamefont
  {Bradbury}, \citenamefont {Frostig}, \citenamefont {Hawkins}, \citenamefont
  {Johnson}, \citenamefont {Katariya}, \citenamefont {Leary}, \citenamefont
  {Maclaurin}, \citenamefont {Necula}, \citenamefont {Paszke}, \citenamefont
  {VanderPlas}, \citenamefont {Wanderman-Milne},\ and\ \citenamefont
  {Zhang}}]{Jax2018}%
  \BibitemOpen
  \bibfield  {author} {\bibinfo {author} {\bibfnamefont {J.}~\bibnamefont
  {Bradbury}}, \bibinfo {author} {\bibfnamefont {R.}~\bibnamefont {Frostig}},
  \bibinfo {author} {\bibfnamefont {P.}~\bibnamefont {Hawkins}}, \bibinfo
  {author} {\bibfnamefont {M.~J.}\ \bibnamefont {Johnson}}, \bibinfo {author}
  {\bibfnamefont {Y.}~\bibnamefont {Katariya}}, \bibinfo {author}
  {\bibfnamefont {C.}~\bibnamefont {Leary}}, \bibinfo {author} {\bibfnamefont
  {D.}~\bibnamefont {Maclaurin}}, \bibinfo {author} {\bibfnamefont
  {G.}~\bibnamefont {Necula}}, \bibinfo {author} {\bibfnamefont
  {A.}~\bibnamefont {Paszke}}, \bibinfo {author} {\bibfnamefont
  {J.}~\bibnamefont {VanderPlas}}, \bibinfo {author} {\bibfnamefont
  {S.}~\bibnamefont {Wanderman-Milne}},\ and\ \bibinfo {author} {\bibfnamefont
  {Q.}~\bibnamefont {Zhang}},\ }\href {http://github.com/jax-ml/jax} {\bibinfo
  {title} {{JAX}: composable transformations of {P}ython+{N}um{P}y programs}}
  (\bibinfo {year} {2018}),\ \bibinfo {note} {software available from
  \url{http://github.com/jax-ml/jax}}\BibitemShut {NoStop}%
\bibitem [{\citenamefont {York}(1999)}]{York1999}%
  \BibitemOpen
  \bibfield  {author} {\bibinfo {author} {\bibfnamefont {J.~W.}\ \bibnamefont
  {York}, \bibfnamefont {Jr.}},\ }\href
  {https://doi.org/10.1103/PhysRevLett.82.1350} {\bibfield  {journal} {\bibinfo
   {journal} {Phys. Rev. Lett.}\ }\textbf {\bibinfo {volume} {82}},\ \bibinfo
  {pages} {1350} (\bibinfo {year} {1999})}\BibitemShut {NoStop}%
\bibitem [{\citenamefont {Ansorg}\ \emph {et~al.}(2004)\citenamefont {Ansorg},
  \citenamefont {Br{\"u}gmann},\ and\ \citenamefont
  {Tichy}}]{AnsorgBruegmannTichy2004}%
  \BibitemOpen
  \bibfield  {author} {\bibinfo {author} {\bibfnamefont {M.}~\bibnamefont
  {Ansorg}}, \bibinfo {author} {\bibfnamefont {B.}~\bibnamefont
  {Br{\"u}gmann}},\ and\ \bibinfo {author} {\bibfnamefont {W.}~\bibnamefont
  {Tichy}},\ }\href {https://doi.org/10.1103/PhysRevD.70.064011} {\bibfield
  {journal} {\bibinfo  {journal} {Phys. Rev. D}\ }\textbf {\bibinfo {volume}
  {70}},\ \bibinfo {pages} {064011} (\bibinfo {year} {2004})}\BibitemShut
  {NoStop}%
\bibitem [{\citenamefont {Smolyak}(1963)}]{Smolyak1963}%
  \BibitemOpen
  \bibfield  {author} {\bibinfo {author} {\bibfnamefont {S.~A.}\ \bibnamefont
  {Smolyak}},\ }\href@noop {} {\bibfield  {journal} {\bibinfo  {journal}
  {Soviet Math. Dokl.}\ }\textbf {\bibinfo {volume} {4}},\ \bibinfo {pages}
  {240} (\bibinfo {year} {1963})},\ \bibinfo {note} {{R}ussian original: Dokl.
  Akad. Nauk SSSR \textbf{148}(5), 1042--1045 (1963)}\BibitemShut {NoStop}%
\bibitem [{\citenamefont {Sirovich}(1987)}]{Sirovich1987}%
  \BibitemOpen
  \bibfield  {author} {\bibinfo {author} {\bibfnamefont {L.}~\bibnamefont
  {Sirovich}},\ }\href {https://doi.org/10.1090/qam/910462} {\bibfield
  {journal} {\bibinfo  {journal} {Quart. Appl. Math.}\ }\textbf {\bibinfo
  {volume} {45}},\ \bibinfo {pages} {561} (\bibinfo {year} {1987})}\BibitemShut
  {NoStop}%
\bibitem [{\citenamefont {Berkooz}\ \emph {et~al.}(1993)\citenamefont
  {Berkooz}, \citenamefont {Holmes},\ and\ \citenamefont
  {Lumley}}]{BerkoozHolmesLumley1993}%
  \BibitemOpen
  \bibfield  {author} {\bibinfo {author} {\bibfnamefont {G.}~\bibnamefont
  {Berkooz}}, \bibinfo {author} {\bibfnamefont {P.}~\bibnamefont {Holmes}},\
  and\ \bibinfo {author} {\bibfnamefont {J.~L.}\ \bibnamefont {Lumley}},\
  }\href {https://doi.org/10.1146/annurev.fl.25.010193.002543} {\bibfield
  {journal} {\bibinfo  {journal} {Annu. Rev. Fluid Mech.}\ }\textbf {\bibinfo
  {volume} {25}},\ \bibinfo {pages} {539} (\bibinfo {year} {1993})}\BibitemShut
  {NoStop}%
\bibitem [{\citenamefont {Husa}\ \emph {et~al.}(2008)\citenamefont {Husa},
  \citenamefont {Hannam}, \citenamefont {Gonz{\'a}lez}, \citenamefont
  {Sperhake},\ and\ \citenamefont {Br{\"u}gmann}}]{Husa2008}%
  \BibitemOpen
  \bibfield  {author} {\bibinfo {author} {\bibfnamefont {S.}~\bibnamefont
  {Husa}}, \bibinfo {author} {\bibfnamefont {M.}~\bibnamefont {Hannam}},
  \bibinfo {author} {\bibfnamefont {J.~A.}\ \bibnamefont {Gonz{\'a}lez}},
  \bibinfo {author} {\bibfnamefont {U.}~\bibnamefont {Sperhake}},\ and\
  \bibinfo {author} {\bibfnamefont {B.}~\bibnamefont {Br{\"u}gmann}},\ }\href
  {https://doi.org/10.1103/PhysRevD.77.044037} {\bibfield  {journal} {\bibinfo
  {journal} {Phys. Rev. D}\ }\textbf {\bibinfo {volume} {77}},\ \bibinfo
  {pages} {044037} (\bibinfo {year} {2008})}\BibitemShut {NoStop}%
\bibitem [{\citenamefont {Walther}\ \emph {et~al.}(2009)\citenamefont
  {Walther}, \citenamefont {Br{\"u}gmann},\ and\ \citenamefont
  {M{\"u}ller}}]{Walther2009}%
  \BibitemOpen
  \bibfield  {author} {\bibinfo {author} {\bibfnamefont {B.}~\bibnamefont
  {Walther}}, \bibinfo {author} {\bibfnamefont {B.}~\bibnamefont
  {Br{\"u}gmann}},\ and\ \bibinfo {author} {\bibfnamefont {D.}~\bibnamefont
  {M{\"u}ller}},\ }\href {https://doi.org/10.1103/PhysRevD.79.124040}
  {\bibfield  {journal} {\bibinfo  {journal} {Phys. Rev. D}\ }\textbf {\bibinfo
  {volume} {79}},\ \bibinfo {pages} {124040} (\bibinfo {year}
  {2009})}\BibitemShut {NoStop}%
\bibitem [{\citenamefont {Ciarfella}\ \emph {et~al.}(2024)\citenamefont
  {Ciarfella}, \citenamefont {Healy}, \citenamefont {Lousto},\ and\
  \citenamefont {Nakano}}]{Healy2024}%
  \BibitemOpen
  \bibfield  {author} {\bibinfo {author} {\bibfnamefont {A.}~\bibnamefont
  {Ciarfella}}, \bibinfo {author} {\bibfnamefont {J.}~\bibnamefont {Healy}},
  \bibinfo {author} {\bibfnamefont {C.~O.}\ \bibnamefont {Lousto}},\ and\
  \bibinfo {author} {\bibfnamefont {H.}~\bibnamefont {Nakano}},\ }\href@noop {}
  {\bibinfo {title} {Quasicircular orbital parameters for numerical relativity
  revisited}} (\bibinfo {year} {2024}),\ \Eprint
  {https://arxiv.org/abs/2406.11564} {arXiv:2406.11564 [gr-qc]} \BibitemShut
  {NoStop}%
\bibitem [{\citenamefont {Peters}(1964)}]{Peters1964}%
  \BibitemOpen
  \bibfield  {author} {\bibinfo {author} {\bibfnamefont {P.~C.}\ \bibnamefont
  {Peters}},\ }\href {https://doi.org/10.1103/PhysRev.136.B1224} {\bibfield
  {journal} {\bibinfo  {journal} {Phys. Rev.}\ }\textbf {\bibinfo {volume}
  {136}},\ \bibinfo {pages} {B1224} (\bibinfo {year} {1964})}\BibitemShut
  {NoStop}%
\bibitem [{\citenamefont {Bernstein}(1912)}]{Bernstein1912}%
  \BibitemOpen
  \bibfield  {author} {\bibinfo {author} {\bibfnamefont {S.}~\bibnamefont
  {Bernstein}},\ }\href@noop {} {\bibfield  {journal} {\bibinfo  {journal}
  {M\'em. Acad. Roy. Belg. (2)}\ }\textbf {\bibinfo {volume} {4}},\ \bibinfo
  {pages} {1} (\bibinfo {year} {1912})}\BibitemShut {NoStop}%
\bibitem [{\citenamefont {Cook}\ and\ \citenamefont
  {York}(1990)}]{CookYork1990}%
  \BibitemOpen
  \bibfield  {author} {\bibinfo {author} {\bibfnamefont {G.~B.}\ \bibnamefont
  {Cook}}\ and\ \bibinfo {author} {\bibfnamefont {J.~W.}\ \bibnamefont {York},
  \bibfnamefont {Jr.}},\ }\href {https://doi.org/10.1103/PhysRevD.41.1077}
  {\bibfield  {journal} {\bibinfo  {journal} {Phys. Rev. D}\ }\textbf {\bibinfo
  {volume} {41}},\ \bibinfo {pages} {1077} (\bibinfo {year}
  {1990})}\BibitemShut {NoStop}%
\bibitem [{\citenamefont {Baumgarte}\ and\ \citenamefont
  {Shapiro}(2010)}]{BaumgarteShapiro2010}%
  \BibitemOpen
  \bibfield  {author} {\bibinfo {author} {\bibfnamefont {T.~W.}\ \bibnamefont
  {Baumgarte}}\ and\ \bibinfo {author} {\bibfnamefont {S.~L.}\ \bibnamefont
  {Shapiro}},\ }\href {https://doi.org/10.1017/CBO9781139193344} {\emph
  {\bibinfo {title} {Numerical Relativity: Solving {E}instein's Equations on
  the Computer}}}\ (\bibinfo  {publisher} {Cambridge University Press},\
  \bibinfo {address} {Cambridge},\ \bibinfo {year} {2010})\BibitemShut
  {NoStop}%
\bibitem [{\citenamefont {Cook}(1994)}]{Cook1994}%
  \BibitemOpen
  \bibfield  {author} {\bibinfo {author} {\bibfnamefont {G.~B.}\ \bibnamefont
  {Cook}},\ }\href {https://doi.org/10.1103/PhysRevD.50.5025} {\bibfield
  {journal} {\bibinfo  {journal} {Phys. Rev. D}\ }\textbf {\bibinfo {volume}
  {50}},\ \bibinfo {pages} {5025} (\bibinfo {year} {1994})}\BibitemShut
  {NoStop}%
\bibitem [{\citenamefont {Lovelace}(2009)}]{Lovelace2009}%
  \BibitemOpen
  \bibfield  {author} {\bibinfo {author} {\bibfnamefont {G.}~\bibnamefont
  {Lovelace}},\ }\href {https://doi.org/10.1088/0264-9381/26/11/114002}
  {\bibfield  {journal} {\bibinfo  {journal} {Class. Quantum Grav.}\ }\textbf
  {\bibinfo {volume} {26}},\ \bibinfo {pages} {114002} (\bibinfo {year}
  {2009})},\ \Eprint {https://arxiv.org/abs/0812.3132} {arXiv:0812.3132
  [gr-qc]} \BibitemShut {NoStop}%
\bibitem [{\citenamefont {Dain}\ \emph {et~al.}(2008)\citenamefont {Dain},
  \citenamefont {Lousto},\ and\ \citenamefont
  {Zlochower}}]{DainLoustoZlochower2008}%
  \BibitemOpen
  \bibfield  {author} {\bibinfo {author} {\bibfnamefont {S.}~\bibnamefont
  {Dain}}, \bibinfo {author} {\bibfnamefont {C.~O.}\ \bibnamefont {Lousto}},\
  and\ \bibinfo {author} {\bibfnamefont {Y.}~\bibnamefont {Zlochower}},\ }\href
  {https://doi.org/10.1103/PhysRevD.78.024039} {\bibfield  {journal} {\bibinfo
  {journal} {Phys. Rev. D}\ }\textbf {\bibinfo {volume} {78}},\ \bibinfo
  {pages} {024039} (\bibinfo {year} {2008})},\ \Eprint
  {https://arxiv.org/abs/0803.0351} {arXiv:0803.0351 [gr-qc]} \BibitemShut
  {NoStop}%
\bibitem [{\citenamefont {Lovelace}\ \emph {et~al.}(2008)\citenamefont
  {Lovelace}, \citenamefont {Owen}, \citenamefont {Pfeiffer},\ and\
  \citenamefont {Chu}}]{Lovelace2008}%
  \BibitemOpen
  \bibfield  {author} {\bibinfo {author} {\bibfnamefont {G.}~\bibnamefont
  {Lovelace}}, \bibinfo {author} {\bibfnamefont {R.}~\bibnamefont {Owen}},
  \bibinfo {author} {\bibfnamefont {H.~P.}\ \bibnamefont {Pfeiffer}},\ and\
  \bibinfo {author} {\bibfnamefont {T.}~\bibnamefont {Chu}},\ }\href
  {https://doi.org/10.1103/PhysRevD.78.084017} {\bibfield  {journal} {\bibinfo
  {journal} {Phys. Rev. D}\ }\textbf {\bibinfo {volume} {78}},\ \bibinfo
  {pages} {084017} (\bibinfo {year} {2008})},\ \Eprint
  {https://arxiv.org/abs/0805.4192} {arXiv:0805.4192 [gr-qc]} \BibitemShut
  {NoStop}%
\bibitem [{\citenamefont {Ruchlin}\ \emph {et~al.}(2018)\citenamefont
  {Ruchlin}, \citenamefont {Etienne},\ and\ \citenamefont
  {Baumgarte}}]{RuchlinEtienneBaumgarte2018}%
  \BibitemOpen
  \bibfield  {author} {\bibinfo {author} {\bibfnamefont {I.}~\bibnamefont
  {Ruchlin}}, \bibinfo {author} {\bibfnamefont {Z.~B.}\ \bibnamefont
  {Etienne}},\ and\ \bibinfo {author} {\bibfnamefont {T.~W.}\ \bibnamefont
  {Baumgarte}},\ }\href {https://doi.org/10.1103/PhysRevD.97.064036} {\bibfield
   {journal} {\bibinfo  {journal} {Phys. Rev. D}\ }\textbf {\bibinfo {volume}
  {97}},\ \bibinfo {pages} {064036} (\bibinfo {year} {2018})},\ \Eprint
  {https://arxiv.org/abs/1712.07658} {arXiv:1712.07658 [gr-qc]} \BibitemShut
  {NoStop}%
\bibitem [{\citenamefont {{The NRPy Developers}}(2026)}]{NRPy}%
  \BibitemOpen
  \bibfield  {author} {\bibinfo {author} {\bibnamefont {{The NRPy
  Developers}}},\ }\href {https://github.com/nrpy/nrpy} {\bibinfo {title}
  {{NRPy}: a {P}ython/{S}ym{P}y-based symbolic code generation toolkit for
  numerical relativity and relativistic astrophysics}} (\bibinfo {year}
  {2026}),\ \bibinfo {note} {software available from
  \url{https://github.com/nrpy/nrpy}; version \texttt{2.2026.6}}\BibitemShut
  {NoStop}%
\bibitem [{\citenamefont {Galassi}\ \emph {et~al.}(2009)\citenamefont {Galassi}
  \emph {et~al.}}]{GSL}%
  \BibitemOpen
  \bibfield  {author} {\bibinfo {author} {\bibfnamefont {M.}~\bibnamefont
  {Galassi}} \emph {et~al.},\ }\href@noop {} {\emph {\bibinfo {title} {{GNU}
  Scientific Library Reference Manual}}},\ \bibinfo {edition} {3rd}\ ed.\
  (\bibinfo  {publisher} {Network Theory Ltd.},\ \bibinfo {year}
  {2009})\BibitemShut {NoStop}%
\bibitem [{\citenamefont {{GRTL Collaboration}}(2026)}]{GRTeclyn}%
  \BibitemOpen
  \bibfield  {author} {\bibinfo {author} {\bibnamefont {{GRTL
  Collaboration}}},\ }\href {https://github.com/GRTLCollaboration/GRTeclyn}
  {\bibinfo {title} {{GRTeclyn}: a code for numerical relativity built on
  {AMReX}}} (\bibinfo {year} {2026}),\ \bibinfo {note} {software available from
  \url{https://github.com/GRTLCollaboration/GRTeclyn}; commit
  \texttt{2bfd19e}}\BibitemShut {NoStop}%
\end{thebibliography}%

\end{document}